\documentclass[
unsortedaddress,
preprint,
 amsmath,amssymb,
 aps,
pre,
]{revtex4-2}

\usepackage{amsmath}
\usepackage{amssymb}
\usepackage{bm}
\usepackage{mathrsfs}
\usepackage{graphicx}
\usepackage{xcolor}
\usepackage{booktabs} 
\usepackage{subfig}
\usepackage{microtype}

\begin{document}

\title{Vortex dynamics and air entrainment in dam break wave impacting on vertical walls: A multiphase lattice Boltzmann study}

\author{Massimiliano Mustè}
\email{mas.muste@stud.uniroma3.it}
\affiliation{Department of Civil, Computer Science and Aeronautical Technologies Engineering,
Roma Tre University, Via Vito Volterra 62, 00146 Rome, Italy}

\author{Andrea Montessori}
\affiliation{Department of Civil, Computer Science and Aeronautical Technologies Engineering,
Roma Tre University, Via Vito Volterra 62, 00146 Rome, Italy}

\author{Pietro Prestininzi}
\affiliation{Department of Civil, Computer Science and Aeronautical Technologies Engineering,
Roma Tre University, Via Vito Volterra 62, 00146 Rome, Italy}

\begin{abstract}

Air entrainment often plays a crucial role in determining impact loads exerted by free-surface wave flows interacting with structures, yet its modelling is often oversimplified in numerical approaches.

In this study a two-phase numerical model, based on the Lattice Boltzmann Method coupled to a conservative Allen–Cahn interface-capturing equation is employed to perform direct numerical simulations of dam-break waves propagating over a dry bed and impacting on vertical walls.

Access to high-resolution simulations enables a detailed assessment of how accurately resolving both air–water and solid–water interfaces affects local and overall dynamics, as well as quantities
of extreme engineering interest. Indeed, the magnitudes and locations of the pressure peaks are strongly affected by wave front deflection and local aeration induced by a small corner vortex.

Additionally, comparisons between no-slip and free-slip implementations suggest that the large air cavity formation, commonly observed as trapped inside the reflected jet falling back onto the incoming flow, may be the result of modeling assumptions rather than intrinsic flow physics, again highlighting the key role of near-wall shear in jet breakup dynamics.  
\end{abstract}

\maketitle









\section{Introduction}

The impact of rapidly moving free-surface flows on vertical structures serves as a conceptual prototype for a broad class of problems in hydraulic and coastal engineering. These flows generate highly transient loads, whose magnitude, duration, and spatial distribution represent critical variables for structural design and risk assessment. Frequently studied scenarios include tsunami bore impacts \cite{xu2021mitigation}, urban flooding against isolated obstacles \cite{la2020discrete}, \cite{kleefsman2005volume}, \cite{aureli2015experimental}, \cite{ozmen2011dam}, bridge piers \cite{adekoya2026combined}, and surge impact on vertical walls \cite{shen2020spatiotemporal}, \cite{liu2022experimental}, \cite{mokrani2016conditions}, \cite{rozki2025physical}, \cite{del2024modelling}, their foundations \cite{huo2023experimental} or in presence of overtopping flow \cite{rozki2025first}.

In these configurations, the engineering response is not determined solely by the hydrodynamic properties of the flow, such as its kinematics \cite{tan2023experimental}, but also by the impact dynamics and the resulting fluid–structure interaction. These aspects govern both the magnitude of impact forces and the spatio-temporal distribution of pressure peaks. In particular, air entrapment and the degree of air–water mixing can drastically influence these pressure peaks, significantly modifying the impulsive loads transmitted to the structure \cite{lugni2006wave}, \cite{bredmose2015violent}.

In this regard, the role of air during the impact of a surge on a vertical wall was investigated by \cite{rozki2025physical}. Through multiphase numerical simulations based on the Reynolds-Averaged 
Navier–Stokes (RANS) equations, the study focused on the second force peak, which was 
attributed to the collapse of the upward water jet onto the incoming flow, trapping a 
macroscopic and well-defined air cavity. During cavity closure, the adopted incompressible simulations, both 2D and 3D, exhibited a negative flow divergence in that area, indicating the onset of compressibility effects. The authors demonstrated that the two configurations predicted this peak very differently: the 2D model produced an unrealistically spurious peak, whereas the 3D model matched the experimental data much better, probably because it was able to better represent the air-escape mechanism, which is intrinsically related to the three-dimensional nature of the phenomenon. Consequently, although the 3D simulation was unable to perfectly reproduce the experimentally observed flow, the simulated air escape mechanism was considered sufficiently representative of the physical process to ensure consistency of the numerical results.

In many numerical simulations, the post-impact dynamics lead to the formation of a large, well-defined air cavity trapped between the reflected jet and the incoming wave \cite{colagrossi2003numerical}, \cite{yang2020numerical}. In this configuration, the entrapped air acts as a compressible cushion capable of generating a nearly uniform pressure peak along the wall. However, experimental observations suggest a more complex physical process, in which no single confined air cavity develops. Instead, a highly turbulent region forms, characterized by an air–water mixture with intense two-phase mixing \cite{tan2023experimental}.

This aspect remains a source of uncertainty in the numerical modelling of dam-break impact flows. In particular, there is currently no clear consensus as to whether the large air cavities constitute an intrinsic feature of the physical phenomenon or, alternatively, arise from modelling assumptions, insufficient spatial resolution in the near-wall region, or simplified treatment of wall boundary conditions.

Despite its apparent geometric simplicity, the impact of a dam-break wave against a vertical wall remains a highly complex multiphase flow problem. A reliable interpretation of the resulting impact loads therefore requires numerical models capable of accurately resolving both the air–water interface and the near-wall flow dynamics.

High-fidelity, interface-resolved numerical simulations provide a valuable route to complement experiments and theoretical models by giving access to hydrodynamic fields that are difficult to measure directly. At the same time, the numerical method must be sufficiently robust to handle violent interface deformation, large density and viscosity ratios, and complex air--water mixing, while preserving the correct influence of solid boundaries.




In this work, we investigate the role of
air entrainment in determining the impact dynamics of dam-break waves on vertical walls by means of high-resolution, three-dimensional multiphase simulations. The numerical framework is based on a thread-safe lattice Boltzmann method for the hydrodynamics, coupled with a conservative Allen--Cahn equation for the liquid--gas in \cite{lauricella2025acclb}, \cite{lauricella2025thread}, \cite{montessori2026breakdown}, \cite{montessori2024high}. The model is applied to benchmark dam-break configurations (\cite{lobovsky2014experimental},\cite{tan2023experimental}) and compared with available experimental measurements of free-surface evolution, wave-front propagation, water levels, pressure signals, and velocity fields.

The simulations are used as a diagnostic tool to clarify the mechanisms governing local impact dynamics and pressure generation. Furthermore, by comparing no-slip and free-slip simulations under otherwise identical conditions, the influence of boundary-layer resolution and near-wall shear on jet breakup, air–water mixing, and the resulting pressure response is assessed. The results indicate that accurate modelling of both the air–water interface and the solid-wall boundary condition is necessary for reliable predictions of dam-break impact loads, and provide useful guidance for high-fidelity simulations of hydraulic and coastal engineering flows.

The structure of the paper is as follows: Section II presents the method, while Section III describes the scaling adopted for the model setup. Section IV reports the benchmark cases and the corresponding results, while the discussion is presented in Section V. Finally, conclusions are drawn in Section VI.

\newpage
\section{Method}
\label{subsec1}

\subsection{Coupled recursive-regularized thread-safe lattice Boltzmann and finite-difference MUSCL scheme}
\label{subsec:lb_cace}

The dam-break simulations are performed with a diffuse-interface multiphase flow solver
based on a hybrid lattice Boltzmann--finite-difference formulation, as implemented in the code accLB (\cite{lauricella2025acclb}, \cite{lauricella2025thread}, \cite{montessori2026breakdown}, \cite{montessori2024high}). The hydrodynamics is
advanced through a variable-density, variable-viscosity lattice Boltzmann method, whereas
the liquid--gas interface is captured by a conservative Allen--Cahn equation discretized with
a finite-difference MUSCL--TVD scheme. The formulation is designed to retain locality at
the algorithmic level, so that the lattice Boltzmann update can be implemented in a
thread-safe manner on shared-memory and GPU architectures, while the phase-field solver
provides conservative and robust interface transport during the large deformations occurring
in dam-break flows.

In the low-Mach-number and small-Knudsen-number limits, the kinetic formulation recovers
the variable-density incompressible Navier--Stokes equations
\begin{equation}
\partial_t \rho + \nabla \cdot (\rho \mathbf{u}) = 0,
\label{eq:continuity}
\end{equation}
\begin{equation}
\partial_t(\rho \mathbf{u}) + \nabla \cdot (\rho \mathbf{u}\otimes\mathbf{u})
=
-\nabla p
+
\nabla \cdot
\left[
\rho \nu
\left(
\nabla \mathbf{u} + \nabla \mathbf{u}^{T}
\right)
\right]
+
\mathbf{F}_{\sigma}
+
\mathbf{F}_{\mathrm{ext}},
\label{eq:momentum}
\end{equation}
where $\rho$ is the local mixture density, $\mathbf{u}$ is the velocity field, $\nu$ is the
kinematic viscosity, and $p$ is the hydrodynamic pressure. The term
$\mathbf{F}_{\sigma}$ denotes the capillary force associated with the diffuse interface,
whereas $\mathbf{F}_{\mathrm{ext}}$ accounts for external body forces. In the present
dam-break configuration, the latter includes gravity,
\begin{equation}
\mathbf{F}_{\mathrm{ext}} = \rho \mathbf{g}.
\label{eq:gravity_force}
\end{equation}

The lattice Boltzmann equation is solved on a $q$-velocity lattice, here a D3Q27 stencil,
with discrete velocities $\{\mathbf{c}_i\}_{i=0}^{q-1}$, weights $\{w_i\}_{i=0}^{q-1}$,
and lattice time step $\Delta t = 1$. The evolution equation reads
\begin{equation}
f_i(\mathbf{x}+\mathbf{c}_i,t+1)
=
f_i(\mathbf{x},t)
+
\omega
\left[
f_i^{\mathrm{eq}}(\mathbf{x},t)-f_i(\mathbf{x},t)
\right]
+
S_i(\mathbf{x},t),
\label{eq:lbe}
\end{equation}
where $f_i$ are the particle distribution functions, $\omega$ is the relaxation frequency,
and $S_i$ is a forcing source term. The kinematic viscosity is related to the relaxation
frequency by
\begin{equation}
\nu =
c_s^2
\left(
\frac{1}{\omega}
-
\frac{1}{2}
\right),
\label{eq:viscosity_lbm}
\end{equation}
with $c_s$ the lattice sound speed.

A key ingredient of the solver is the thread-safe reconstruction of both equilibrium and
non-equilibrium populations from local macroscopic quantities. In this formulation, each
lattice node can be updated independently, avoiding concurrent read--write operations on
neighbouring populations during streaming and collision. This single-pass update is
particularly suited to massively parallel implementations and avoids race conditions on
shared-memory architectures.

We employ an incompressible-pressure formulation in which the zeroth kinetic moment gives
a dimensionless pressure-like variable $p^{*}$. The physical pressure is defined as
\begin{equation}
p = \rho c_s^2 p^{*}.
\label{eq:pressure}
\end{equation}
The macroscopic fields are recovered as
\begin{equation}
p^{*}(\mathbf{x},t) = \sum_i f_i(\mathbf{x},t),
\label{eq:pstar}
\end{equation}
\begin{equation}
\mathbf{u}(\mathbf{x},t)
=
\sum_i f_i(\mathbf{x},t)\mathbf{c}_i
+
\frac{1}{2}
\sum_i S_i(\mathbf{x},t)\mathbf{c}_i .
\label{eq:velocity}
\end{equation}
The density and viscosity are functions of the phase field $\phi \in [0,1]$, with
$\phi=1$ denoting the liquid phase and $\phi=0$ the gas phase. The mixture density is
interpolated as
\begin{equation}
\rho(\phi) = \rho_g + (\rho_l-\rho_g)\phi,
\label{eq:rho_interp}
\end{equation}
whereas the dynamic viscosity is taken as
\begin{equation}
\mu(\phi) = \mu_g + (\mu_l-\mu_g)\phi
\label{eq:mu_interp}
\end{equation}
where subscripts $l$ and $g$ are used to refer to liquid and gas respectively.
The local kinematic viscosity then follows from
\begin{equation}
\nu(\phi) = \frac{\mu(\phi)}{\rho(\phi)}
=
\frac{\mu_g + (\mu_l-\mu_g)\phi}{\rho(\phi)}.
\label{eq:nu_interp}
\end{equation}
Consequently, the relaxation frequency is space-dependent and is computed as
\begin{equation}
\omega(\phi)
=
\left(
\frac{1}{2}
+
\frac{\nu(\phi)}{c_s^2}
\right)^{-1}.
\label{eq:omega_phi}
\end{equation}

The equilibrium distributions are constructed from a third-order Hermite expansion in the
Mach number,
\begin{equation}
f_i^{\mathrm{eq}}
=
w_i
\left[
p^{*}
+
\frac{\mathbf{c}_i\cdot\mathbf{u}}{c_s^2}
+
\frac{\mathbf{H}_i^{(2)}:\mathbf{u}\mathbf{u}}{2c_s^4}
+
\frac{\mathbf{H}_i^{(3)}\vdots\mathbf{u}\mathbf{u}\mathbf{u}}{6c_s^6}
\right],
\label{eq:feq}
\end{equation}
where ``$:$'' denotes double contraction and ``$\vdots$'' denotes triple contraction.
The second- and third-order Hermite tensors are
\begin{equation}
\mathbf{H}_i^{(2)}
=
\mathbf{c}_i\mathbf{c}_i
-
c_s^2 \mathbf{I},
\label{eq:H2}
\end{equation}
\begin{equation}
\mathbf{H}_i^{(3)}
=
\mathbf{c}_i\mathbf{c}_i\mathbf{c}_i
-
c_s^2 \mathcal{S}(\mathbf{c}_i\mathbf{I}),
\label{eq:H3}
\end{equation}
where $\mathbf{I}$ is the identity tensor and $\mathcal{S}(\cdot)$ denotes full
symmetrization over tensor indices.

The non-equilibrium contribution is regularized by projecting it onto the Hermite subspace
up to third order. The populations are decomposed as
\begin{equation}
f_i = f_i^{\mathrm{eq}} + f_i^{\mathrm{neq}},
\end{equation}
with
\begin{equation}
f_i^{\mathrm{neq}}
=
w_i
\left[
\frac{\mathbf{H}_i^{(2)}:\mathbf{A}_{\mathrm{neq}}^{(2)}}{2c_s^4}
+
\frac{\mathbf{H}_i^{(3)}\vdots\mathbf{A}_{\mathrm{neq}}^{(3)}}{6c_s^6}
\right].
\label{eq:fneq}
\end{equation}
At the Navier--Stokes level, the second-order non-equilibrium tensor can be written in
gradient form as
\begin{equation}
\mathbf{A}_{\mathrm{neq}}^{(2)}
=
-
\frac{c_s^2}{\omega }
\left[
\nabla(\rho\mathbf{u})
+
\nabla(\rho\mathbf{u})^{T}
\right],
\label{eq:A2neq}
\end{equation}
whereas the third-order contribution is reconstructed recursively as
\begin{equation}
\mathbf{A}_{\mathrm{neq}}^{(3)}
=
\mathcal{S}
\left(
\mathbf{u}\otimes\mathbf{A}_{\mathrm{neq}}^{(2)}
\right).
\label{eq:A3neq}
\end{equation}
This recursive regularization filters non-hydrodynamic higher-order modes and improves the
robustness of the solver in low-viscosity and high-Reynolds-number regimes.

Body forces are incorporated through the Guo forcing scheme,
\begin{equation}
S_i
=
w_i
\left[
\frac{\mathbf{c}_i-\mathbf{u}}{c_s^2}
+
\frac{(\mathbf{c}_i\cdot\mathbf{u})\mathbf{c}_i}{c_s^4}
\right]
\cdot \mathbf{F},
\label{eq:guo_force}
\end{equation}
and are consistently treated with a trapezoidal time integration by shifting the equilibrium
populations in the collision step. For the present multiphase formulation, the total force is
written as
\begin{equation}
\mathbf{F}
=
\mathbf{F}_{\sigma}
+
\mathbf{F}_{p}
+
\mathbf{F}_{\nu}
+
\mathbf{F}_{\mathrm{ext}}.
\label{eq:total_force}
\end{equation}
The capillary contribution is expressed in chemical-potential form,
\begin{equation}
\mathbf{F}_{\sigma}
=
\mu_{\phi}\nabla\phi,
\label{eq:capillary_force}
\end{equation}
where $\mu_{\phi}$ is the chemical potential associated with the phase-field free energy.
In the present implementation, it is computed as
\begin{equation}
\mu_{\phi}
=
4\beta \phi(\phi-1)
\left(\phi-\frac{1}{2}\right)
-
\kappa_{\phi}\nabla^2\phi,
\label{eq:chemical_potential}
\end{equation}
where $\beta$ and $\kappa_{\phi}$ control the height of the double-well potential and the
interfacial energy penalty, respectively.

Since the pressure is defined as $p=\rho c_s^2 p^{*}$, its gradient can be decomposed as
\begin{equation}
\nabla p
=
\rho c_s^2 \nabla p^{*}
+
p^{*} c_s^2 \nabla \rho .
\label{eq:pressure_split}
\end{equation}
The first term is embedded in the equilibrium populations, whereas the second is introduced
explicitly through the correction force
\begin{equation}
\mathbf{F}_{p}
=
-
p^{*}c_s^2\nabla\rho .
\label{eq:pressure_correction}
\end{equation}
An additional correction is included to recover the variable-coefficient viscous operator in
Eq.~\eqref{eq:momentum}. In compact tensor notation, this correction reads
\begin{equation}
\mathbf{F}_{\nu}
=
-
\frac{\nu\omega}{c_s^2}
\left[
\sum_i
\left(
f_i-f_i^{\mathrm{eq}}
\right)
\mathbf{c}_i\mathbf{c}_i
\right]
\cdot
\nabla\rho .
\label{eq:viscous_correction}
\end{equation}
Here the dot denotes contraction of the second-order tensor with the density-gradient
vector.

The gradients and Laplacians required by the forcing terms and by the phase-field equation
are evaluated using isotropic lattice-consistent finite-difference stencils,
\begin{equation}
\nabla \Psi(\mathbf{x})
=
\frac{1}{c_s^2}
\sum_i
w_i
\Psi(\mathbf{x}+\mathbf{c}_i)
\mathbf{c}_i,
\label{eq:grad_iso}
\end{equation}
\begin{equation}
\nabla^2 \Psi(\mathbf{x})
=
\frac{2}{c_s^2}
\left[
\sum_{i\neq 0}
w_i
\Psi(\mathbf{x}+\mathbf{c}_i)
-
(1-w_0)
\Psi(\mathbf{x})
\right],
\label{eq:lap_iso}
\end{equation}
where $\Psi$ is a generic scalar field.

\subsection{Conservative Allen--Cahn equation and MUSCL--TVD discretization}
\label{subsec:cace_muscl}

The liquid--gas interface is represented by the phase field $\phi$, with $\phi=1$ in the
liquid, $\phi=0$ in the gas, and $\phi=1/2$ identifying the centre of the diffuse interface.
The phase field evolves according to the conservative Allen--Cahn equation,
\begin{equation}
\partial_t \phi
+
\mathbf{u}\cdot\nabla\phi
=
D\nabla^2\phi
-
\gamma
\nabla\cdot
\left[
\phi(1-\phi)\mathbf{n}
\right],
\label{eq:cace_nonconservative}
\end{equation}
where $D$ is the interface diffusivity and
\begin{equation}
\mathbf{n}
=
\frac{\nabla\phi}{|\nabla\phi|}
\label{eq:normal_phi}
\end{equation}
is the interface normal. The last term in Eq.~\eqref{eq:cace_nonconservative} is a
compressive anti-diffusive flux that counteracts numerical diffusion and maintains a smooth
interface of prescribed thickness $\delta$. The compression coefficient is chosen as
\begin{equation}
\gamma = \frac{4D}{\delta}.
\label{eq:gamma_cace}
\end{equation}
For improved conservation at the discrete level, the advective term is written in conservative
form. Introducing the advective flux
\begin{equation}
\mathbf{j}_{a} = \mathbf{u}\phi,
\end{equation}
the phase-field equation is advanced as
\begin{equation}
\partial_t \phi
+
\nabla\cdot\mathbf{j}_{a}
=
D\nabla^2\phi
-
\gamma
\nabla\cdot
\left[
\phi(1-\phi)\mathbf{n}
\right].
\label{eq:cace_conservative}
\end{equation}

In lattice units, the update at a fluid node $(i,j,k)$ is written as
\begin{equation}
\phi_{ijk}^{n+1}
=
\phi_{ijk}^{n}
-
(\nabla_h\cdot\mathbf{j}_{a})_{ijk}
+
\tau_{\mathrm{diff}}
(\nabla_h^2\phi)_{ijk}
+
S_{ijk}^{\mathrm{comp}},
\label{eq:phi_update}
\end{equation}
where $\nabla_h$ and $\nabla_h^2$ denote discrete Cartesian operators,
$\tau_{\mathrm{diff}}$ contains the coefficients associated with the diffusive term, and
$S_{ijk}^{\mathrm{comp}}$ is the discrete counterpart of the compressive flux. A
double-buffering strategy is used for the phase field, and the update is applied only to
fluid nodes.

The advective flux is discretized by a three-dimensional MUSCL--TVD scheme with minmod
limiting. Assuming $\Delta x=\Delta y=\Delta z=1$, the discrete divergence of the advective
flux is
\begin{align}
(\nabla_h\cdot\mathbf{j}_{a})_{ijk}
=&
\left(
j^{x}_{i+1/2,j,k}
-
j^{x}_{i-1/2,j,k}
\right)
+
\left(
j^{y}_{i,j+1/2,k}
-
j^{y}_{i,j-1/2,k}
\right)
\nonumber\\
&+
\left(
j^{z}_{i,j,k+1/2}
-
j^{z}_{i,j,k-1/2}
\right),
\label{eq:flux_divergence}
\end{align}
where
\begin{equation}
j^{x}_{i+1/2,j,k}
=
u_{i+1/2,j,k}
\phi^{\mathrm{up}}_{i+1/2,j,k},
\quad
j^{y}_{i,j+1/2,k}
=
v_{i,j+1/2,k}
\phi^{\mathrm{up}}_{i,j+1/2,k},
\quad
j^{z}_{i,j,k+1/2}
=
w_{i,j,k+1/2}
\phi^{\mathrm{up}}_{i,j,k+1/2}.
\label{eq:face_fluxes}
\end{equation}
The face-centred velocities are obtained by arithmetic interpolation of neighbouring
cell-centred values, e.g.
\begin{equation}
u_{i+1/2,j,k}
=
\frac{1}{2}
\left(
u_{i,j,k}
+
u_{i+1,j,k}
\right),
\label{eq:face_velocity}
\end{equation}
with analogous expressions for the other velocity components.

The upwind value of $\phi$ at each face is obtained by a piecewise-linear MUSCL
reconstruction. Along the $x$ direction, the limited slope is
\begin{equation}
s^{x}_{i,j,k}
=
\mathrm{minmod}
\left(
\phi_{i,j,k}-\phi_{i-1,j,k},
\phi_{i+1,j,k}-\phi_{i,j,k}
\right),
\label{eq:minmod_slope}
\end{equation}
where
\begin{equation}
\mathrm{minmod}(a,b)
=
\begin{cases}
0, & ab \leq 0, \\
\mathrm{sign}(a)\min(|a|,|b|), & ab>0.
\end{cases}
\label{eq:minmod}
\end{equation}
The left and right reconstructed states at the face $i+1/2$ are
\begin{equation}
\phi^{L}_{i+1/2,j,k}
=
\phi_{i,j,k}
+
\frac{1}{2}s^{x}_{i,j,k},
\qquad
\phi^{R}_{i+1/2,j,k}
=
\phi_{i+1,j,k}
-
\frac{1}{2}s^{x}_{i+1,j,k}.
\label{eq:left_right_states}
\end{equation}
The upwind state is then selected as
\begin{equation}
\phi^{\mathrm{up}}_{i+1/2,j,k}
=
\begin{cases}
\phi^{L}_{i+1/2,j,k}, & u_{i+1/2,j,k}\geq 0, \\
\phi^{R}_{i+1/2,j,k}, & u_{i+1/2,j,k}<0.
\end{cases}
\label{eq:upwind_state}
\end{equation}
The same reconstruction is applied in the $y$ and $z$ directions. The resulting scheme is
second-order accurate in smooth regions and locally reverts to first order near steep
gradients, thereby suppressing spurious oscillations at the liquid--gas interface during the
rapid deformation and impact stages of the dam-break evolution.

The compressive contribution is written as the divergence of the flux
\begin{equation}
\mathbf{j}_{c}
=
\phi(1-\phi)\mathbf{n}.
\label{eq:compressive_flux}
\end{equation}
Its discrete contribution is evaluated as
\begin{equation}
S_{ijk}^{\mathrm{comp}}
=
-
\gamma
(\nabla_h\cdot\mathbf{j}_{c})_{ijk}.
\label{eq:compressive_source}
\end{equation}
In the implementation, the divergence of $\mathbf{j}_{c}$ is computed using an isotropic
D3Q27-based stencil, consistent with the lattice operators used in the hydrodynamic solver.
This choice reduces grid-orientation effects and helps preserve a regular diffuse-interface
profile even under the strong stretching and folding induced by the dam-break flow.

\subsection{Boundary conditions}

Boundary conditions are imposed through a non-equilibrium extrapolation approach, following the idea originally proposed by Guo et al. \cite{zhao2002non} and combined here with a regularized reconstruction of the non-equilibrium populations based on Hermite projection \cite{chikatamarla2006grad}. For each boundary node $x_B$, the unknown incoming populations are reconstructed by separating equilibrium and non-equilibrium contributions. The equilibrium part is computed from the macroscopic variables prescribed at the boundary, whereas the non-equilibrium part is extrapolated from the nearest fluid node $x_F$. In compact form, the reconstructed populations can be written as
\begin{equation}
f_i(x_B+\mathbf{c}_i,t+1)=
f_i^{eq}(\rho_B,\mathbf{u}_B)
+
(1-\omega) f_i^{neq,reg}(x_F,t),
\end{equation}
where $f_i^{neq,reg}$ denotes the regularized non-equilibrium contribution obtained from the Hermite reconstruction of the momentum-flux tensor at the adjacent fluid node. This formulation provides a unified way to impose both no-slip and free-slip walls. For no-slip boundaries, the wall velocity is prescribed directly, $\mathbf{u}_B=\mathbf{0}$ for stationary walls, while the density is extrapolated from the neighboring fluid node. For free-slip boundaries, the impermeability condition is enforced by setting the wall-normal velocity to zero, whereas the tangential velocity components are extrapolated from the adjacent fluid node, consistently with a zero-normal-gradient condition. The reconstructed populations are then streamed back into the computational domain, yielding an explicit and memory-efficient boundary treatment fully compatible with the thread-safe LB implementation adopted in the present solver.

\section{Model setup}
\label{sec1}

\subsection{Scaling}
\label{sec:scaling}

All simulations are performed in lattice units \cite{latt2008choice}, \cite{kruger2017lattice}, 
with the latter denoted with a subscript $n$ in the following. For gravity-driven
free-surface flows, such as dam-break impacts, the leading-order balance is
between inertia and gravity. 
The relevant dimensionless number is therefore
the Froude number 
\cite{colagrossi2003numerical}, \cite{stansby1998initial},
\cite{lauber1998experiments}.

\begin{equation}
Fr = \frac{U}{\sqrt{gH}}
\end{equation}
where \(U\) is the characteristic velocity, \(g\) is the gravitational
acceleration, and \(H\) is the initial water depth. In the present scaling,
the characteristic velocity is chosen as the gravity-wave velocity \(Fr=1\),
namely:
\begin{equation}
U\equiv\sqrt{gH}
\label{eq:vel_scale}
\end{equation}
The Mach number, defined as $Ma=U/c_s$, needs to be kept sufficiently small 
to minimize compressibility effects \cite{kruger2017lattice}.
Plugging the Mach number definition into Eq. (\ref{eq:vel_scale}) the numerical
initial depth can be defined:
\begin{equation}
H_n = \frac{Ma^2 c_s^2}{g_n},
\end{equation}
provided the numerical gravitational acceleration $g_n$
is chosen on the order of $10^{-6}$--$10^{-7}$ to ensure stability.
Once $H_n$ is determined, the size of the numerical domain can be 
defined by enforcing geometric similitude with the experimental setup.
Strict Reynolds-number similarity with the experiments is not enforced.
This choice does not conflict with the direct numerical character of the
simulations: here, Direct Numerical Simulation (DNS) denotes that the variable-density incompressible
Navier--Stokes equations coupled to the interface-capturing equation are
solved without any turbulence closure or sub-grid model, and that the
dynamically relevant scales associated with the selected numerical Reynolds
number are explicitly resolved. The simulations should therefore be regarded
as Froude-scaled, interface-resolved DNS at finite Reynolds number, rather
than as exact-Re replicas of the laboratory flow.
The Reynolds number is nevertheless kept sufficiently high for the flow to
remain inertia-dominated and strongly unsteady \cite{heller2011scale}, \cite{heller2017self}. 
Under these conditions,
the main kinematic and impact observables considered in this work---front
propagation, run-up and run-down dynamics, pressure peaks, near-wall air
entrapment, and bulk air--water mixing---are primarily controlled by
inertial--gravitational dynamics and by the correct treatment of the wall
boundary condition. The reduced Reynolds number may affect the extent of
the turbulent cascade, the smallest resolved structures, and the detailed
spectral content of the flow. These aspects are not the focus of the present
study, which does not attempt to characterize turbulence spectra or
small-scale turbulent statistics. Instead, the objective is to assess how
interface resolution and near-wall boundary conditions control the
macroscopic and localized observables relevant to dam-break impact loads.
Once the target Reynolds number is selected,
the lattice kinematic viscosity is obtained from:
\begin{equation}
\nu_n = \frac{\sqrt{g_n H_n^3}}{Re}
\end{equation}
and the corresponding relaxation time is:
\begin{equation}
\tau = \frac{1}{2} + \frac{\nu_n}{c_s^2}
\end{equation}

The resulting parameters and dimensions of the numerical setups are summarized in Table~\ref{tab1}. 
The simulations are performed on a computational grid of approximately $10^8$ 
lattice nodes.

All simulations were performed on the Leonardo supercomputer, using a single 
compute node equipped with four NVIDIA A100 GPUs. A detailed assessment of accLB's performance across both NVIDIA and AMD-based architectures, for single- and multi-component fluids, is reported in \cite{lauricella2025acclb}.

\begin{table}
\centering
\caption{Simulation parameters and numerical setup: NS and FS stand for No-Slip and Free-Slip Boundary Conditions (BC).}
\label{tab1}

\begin{tabular}{lccccccc}
\toprule
Case & Domain (lu) & $\Delta x$ (cm) & $g_n$ & $\tau$ & $Re$ & $Ma$ & BC \\
\midrule
I   &$ 1786\times166 \times 666$ & 0.09 & $1 \times 10^{-6}$ & 0.503 & $6 \times10^3$ & $3.2 \times10^{-2}$ & NS \\
II  &$ 1050 \times 292 \times 700$ & 0.086 & $5 \times 10^{-7}$ & 0.5015 & $5 \times10^3$ & $1.9 \times10^{-2}$ & NS \\
III (a) &$ 1050 \times 292 \times 700$ & 0.086 & $1 \times 10^{-7}$ & 0.5015 & $2.2 \times10^3$ & $8.4 \times10^{-3}$  & NS \\
III (b)  &$ 1050 \times 292 \times 700$ & 0.086 & $1 \times 10^{-7}$ & 0.5015 & $2.2 \times10^3$ & $8.4 \times10^{-3}$   & FS \\
\bottomrule
\end{tabular}
\end{table}

\section{Results}
\label{sec1}

Two case studies are presented in the following, in which the multiphase thread-safe lattice Boltzmann (TSLB) model is benchmarked against experimental data of dam-break flows. These simulations are instrumental in investigating whether the high-resolution adopted in the numerical setup is required to accurately predict local impact dynamics and free-surface wave interactions with vertical structures.

\subsection{Dam-break flow against a vertical wall (Case I)}
\label{subsecA}

\begin{figure}[t]
\centering

\includegraphics[trim={1.0cm 7.0cm 3.5cm 2.5cm}, clip, width=0.9\textwidth]{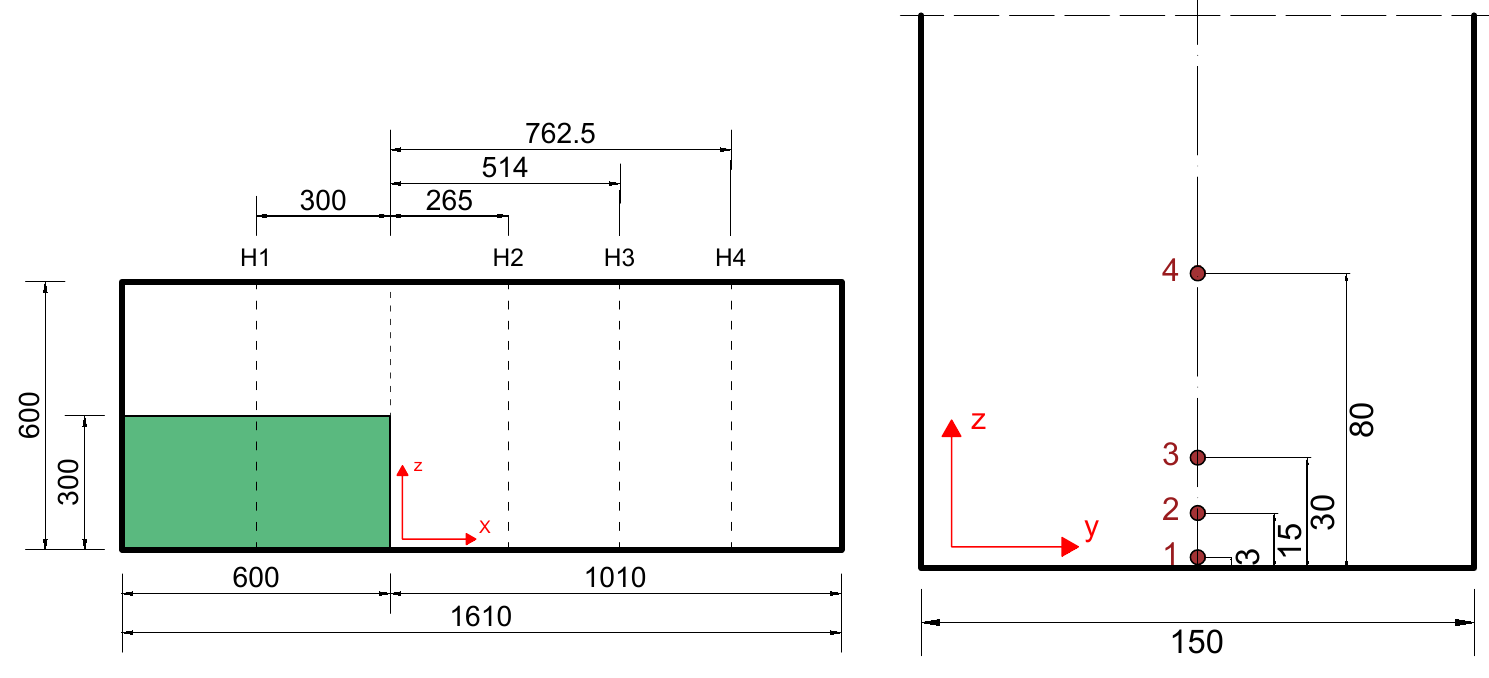}

\caption{Schematic representation of the experimental domain (Case I). Side view of the tank shows the locations of the free-surface level measurements (left) and front view indicating the positions of the pressure sensors along the central axis of the impact wall (right). Dimensions are in millimeters \cite{lobovsky2014experimental}.}

\label{fig:schema}
\end{figure}

The first case study considers the experimental investigation of a dam-break flow over a dry bed impacting a vertical wall, conducted and reported by \cite{lobovsky2014experimental} and shown in Fig.~\ref{fig:schema}. 
The experimental domain has a length of 1.61~m, a height of 0.60~m, and a width of 0.15~m. The initial water column has a height of 0.30~m and a length of 0.60~m. 
Downstream of the gate, pressure sensors are installed on the vertical impact wall at different elevations (3~mm, 15~mm, 30~mm, and 80~mm above the bottom).
The comparison with experimental results is carried out using non-dimensional time $T=t\sqrt{g/H}$, where $t$ is the dimensional time measured from the start of the simulation and $g$ is the gravitational acceleration.

\subsubsection{Free-surface profile}

\begin{figure}
\centering

\includegraphics[trim={0cm 9.6cm 0cm 0cm},clip,width=0.9\textwidth]{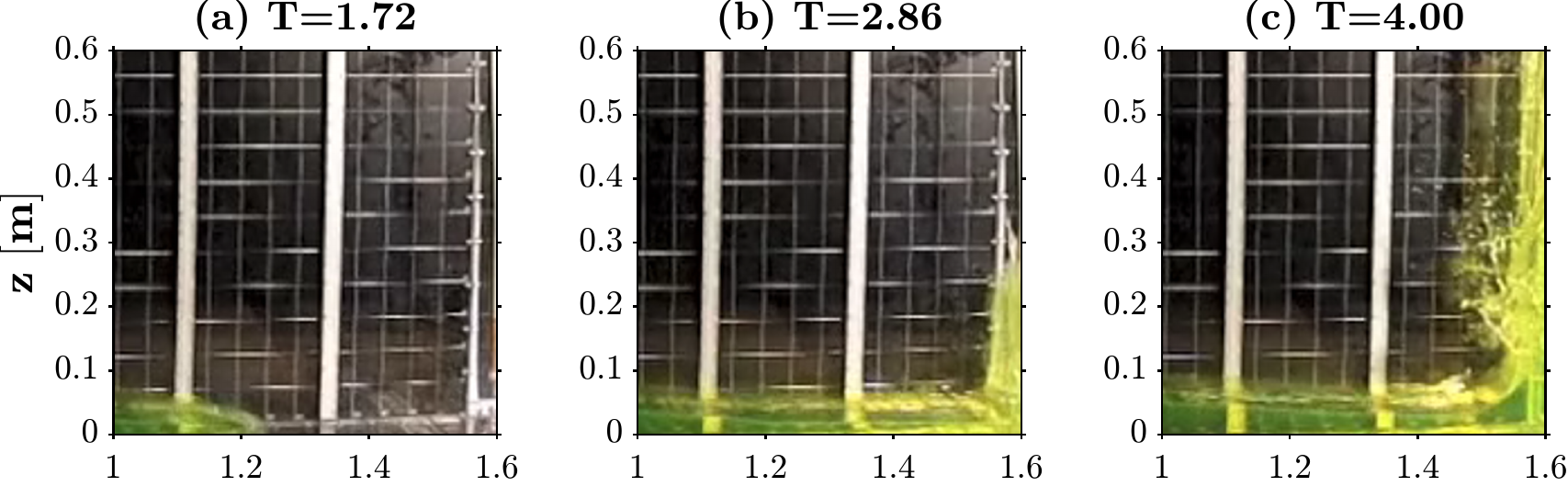}

\vspace{0.2cm}

\includegraphics[trim={0cm 0cm 0cm 9.6cm},clip,width=0.9\textwidth]{figures/SupLib_Lob_exp.pdf}

\vspace{0.4cm}

\includegraphics[trim={0cm 10.4cm 0cm 0cm},clip,width=0.9\textwidth]{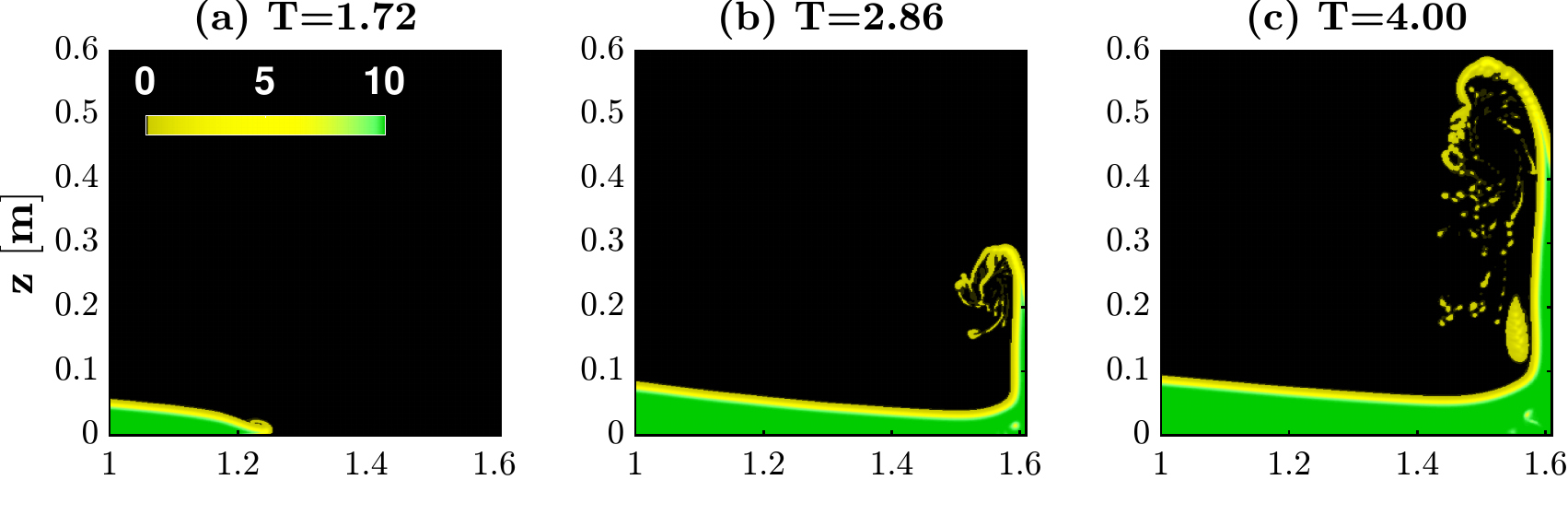}

\vspace{0.2cm}

\includegraphics[trim={0cm 0cm 0cm 9.9cm},clip,width=0.9\textwidth]{figures/SupLib_Lob_num.pdf}

\caption{Comparison between snapshots of the experiments by~\cite{lobovsky2014experimental} (top) and the corresponding numerical density fields in lattice units (bottom). The numerical density ranges from 0.02 (air) to 10 (water).}

\label{fig:profile}
\end{figure}

The qualitative comparison of the free-surface profiles between the experimental test and the numerical model is reported in Fig.~\ref{fig:profile} for different time instants.  

\begin{figure}
\centering

\includegraphics[trim={0cm 11.8cm 0cm 0cm},clip,width=0.6\textwidth]{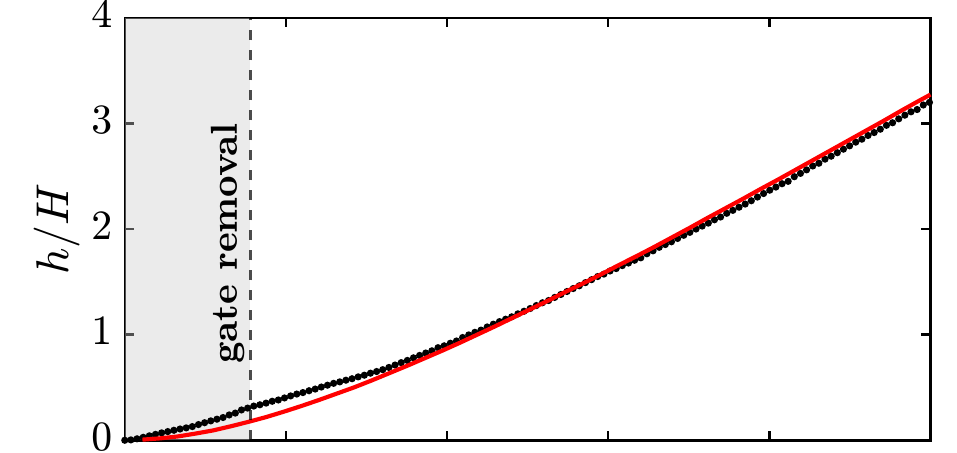}

\includegraphics[trim={0cm 0cm 0cm 9.5cm},clip,width=0.6\textwidth]{figures/FronteOnda_benchmark_Lob.pdf}

\caption{Temporal evolution of the wave front position (top) and velocity (bottom). Comparison between experimental results \cite{lobovsky2014experimental} and TSLB numerical simulation.}

\label{fig:front}
\end{figure}

Both the propagation and the run-up phases are reproduced remarkably well, as shown in the first snapshots of Fig.~\ref{fig:profile}, as well as the plots in Fig.~\ref{fig:front}. Although the height of the tip of the simulated rising jet is lower than the experimental one, the collapse of such column of water, as well as its consequent recession swash phase are simulated in details, with consistent accuracy in the predictions of the water level.

At $T=1.72$ and $T=2.86$, the approach phase of the wave front (primary wave) and the jet run-up along the wall are satisfactorily reproduced. At $T=4.00$, it can also be observed that the jet breaking, with the consequent formation of an aerated mixture, is located slightly above the region where the flow reverses its direction, in agreement with experimental observations.

At $T=5.15$ and $T=6.29$, the run-down process of the main jet takes place. The maximum heights reached by the falling jet impacting the incoming flow are comparable with the experimental ones, as well as the dynamics and the extent of mixing, which appear to be adequately reproduced. In this phase, the air fraction in the mixture decreases on average from $62\%$ (at $T=5.15$) to $47\%$ (at $T=6.29$).

Finally, at $T=7.43$, the return flow (secondary wave) is observed. Compared to the primary wave, it exhibits a higher level of mixing above the free surface, accompanied by significant vorticity generation. This behavior is consistent with the experimental findings of \cite{lobovsky2014experimental} and the numerical simulations of \cite{lubin2003fully}, \cite{lubin2006three}, \cite{landrini2007gridless}.

\subsubsection{Wave front evolution}

The temporal evolution of the wave front position and velocity prior to impact is reported in Fig.~\ref{fig:front}. The numerical front position is assumed to be the largest x location of the average iso-density in the domain.

The velocities of both the experimental and the numerical wave fronts, extracted at the symmetry plane, are computed using a centered finite-difference scheme of the position. A Savitzky--Golay filter (window length 9, polynomial order 3) is applied to both in order to reduce local fluctuations introduced by numerical differentiation, without altering the physical trends of the front propagation.

The main discrepancies between experimental and numerical results are observed at early times and can be attributed to the different initial conditions between the experiment and the simulation, the latter accounting for an istantaneous collapse of the water column, while the former comprises a gradual, though fast, gate removal. When the gate is opened, a high-velocity water jet is released from the gap beneath it, while the remaining water column is still partially obstructed by the lifting gate. Once the gate is fully open, at approximately $t = 0.069\ \mathrm{s}$ \cite{lobovsky2014experimental}, the wave velocity temporarily decreases before increasing again steadily, in agreement with the numerical simulation. This result is consistent with the observations reported by \cite{kamra2018numerical}.

The numerical and experimental wave fronts impinge onto the vertical wall almost simultaneously with very close velocities, roughly $v \approx 1.7\sqrt{gH}$. Such value is lower than the theoretical solution of \cite{ritter1892fortpflanzung} due bed frictional effects.

The comparison of the dimensionless propagation and impact times shows good agreement between numerical simulation and experiment. This confirms that Froude similarity is properly satisfied, and therefore the inertial--gravitational dynamics are adequately captured by the model.

\subsubsection{Water level time histories}

\begin{figure}
\centering

\includegraphics[trim={0cm 11.8cm 17.9cm 0cm},clip,height=0.245\textwidth]{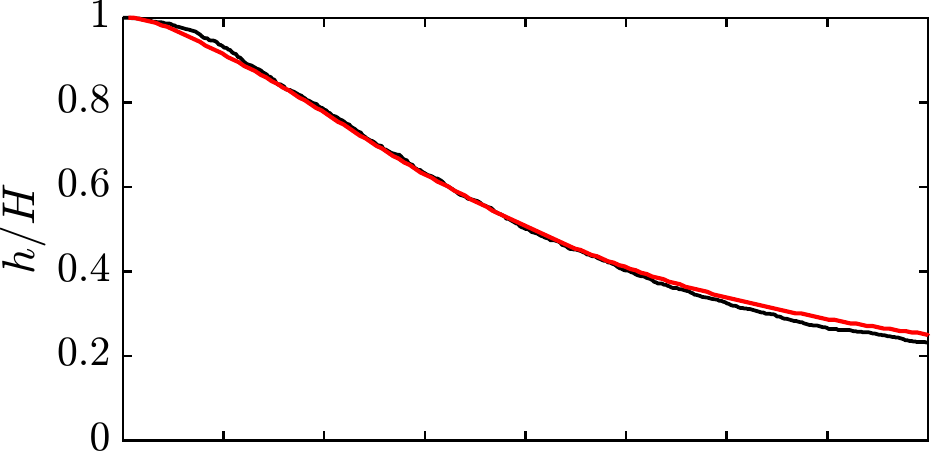}
\includegraphics[trim={19.8cm 11.8cm 0cm 0cm},clip,height=0.245\textwidth]{figures/H_benchmark_Lob.pdf}
\includegraphics[trim={0cm 0cm 17.9cm 9.8cm},clip,height=0.31\textwidth]{figures/H_benchmark_Lob.pdf}
\includegraphics[trim={19.8cm 0cm 0cm 9.8cm},clip,height=0.31\textwidth]{figures/H_benchmark_Lob.pdf}

\caption{Temporal evolution of water levels at sections H1 (top left), H2 (top right), H3 (bottom left), and H4 (bottom right). Comparison between experimental data \cite{lobovsky2014experimental} and TSLB numerical solution.}

\label{fig:level}
\end{figure}

Fig.~\ref{fig:level} shows the temporal evolution of the water levels obtained from the TSLB numerical modeling, compared with the experimental measurements reported by \cite{lobovsky2014experimental}.

The analysis is carried out at the vertical sections H1, H2, H3, and H4, located within the computational domain (as shown in Fig.~\ref{fig:schema}), considering an initial water depth of $H = 0.3\ \mathrm{m}$.

The numerical air-water interface is assumed to be described by a given iso-density surface. The water level is the distance between such interface and the bed. 

At section H1, a progressive decrease in the water level is observed until the end of the simulation. During this interval, the secondary wave generated after the impact does not reach the location of section H1. To track the water level drop due to the downstream propagation of the primary wave, the average numerical density (namely $\rho=5$) is adopted. 

At sections H2, H3, and H4, two distinct phases of motion can be identified. The first is characterized by the downstream propagation of the primary wave before impact with the vertical wall. The second corresponds to the return of the secondary wave after impact. During this return phase, the experimentally measured water levels are higher than in the initial phase. This is due to the secondary wave exhibiting strong mixing between the two phases and no clearly identifiable air-water interface. Such dynamics does not characterize the primary wave, which maintains a more regular profile.

Consequently, in order to carry out a fair comparison with the experimental results, a lower numerical density threshold is required for the secondary wave, as the water near the surface is strongly mixed with air. Using a threshold value of $0.3$ allows the return wave to be adequately represented, whereas a threshold of $5$ is more suitable for describing the water levels associated with the primary wave.

\begin{figure}
\centering
\includegraphics[trim={0.0cm 0.0cm 0.0cm 0.0cm}, clip, width=0.9\textwidth]{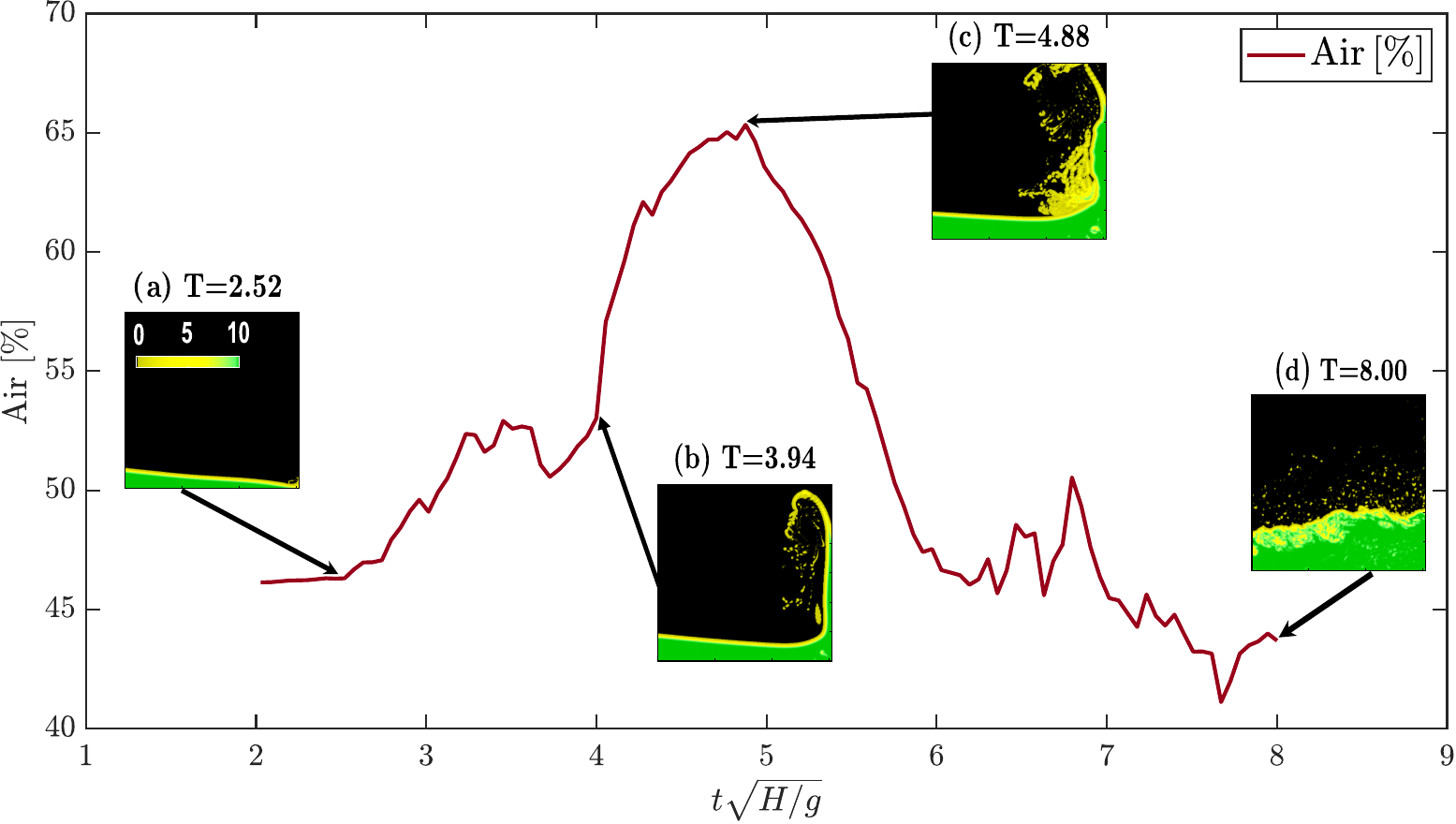}
\caption{Temporal evolution of the air fraction in the mixture. Aeration process during run-up phase and deaeration process during run-down phase, with comparable duration of the rising and falling limbs. Maximum aeration occurs at time $T = 4.88$, corresponding to the complete formation of the counterclockwise vortex.}
\label{fig:mixing}
\end{figure}

\subsubsection{Air fraction in the mixture}

As previously discussed, the TSLB numerical model effectively simulates air--water mixing during a dam-break impact against a vertical wall. This allows for an accurate assessment of the air fraction entrained in the jet throughout the event.

A computational node is classified as belonging to the air–water mixture phase if its density does not coincide with either of the two extreme values, i.e., if it satisfies the condition \(0.02 < \rho_{\mathrm{mix}} < 10\). 

At each time instant, an average value $\bar{\rho}_{mix}$ of the $\rho_{mix}$ distribution in the whole domain is computed, and the volumetric air fraction is evaluated as:
\begin{equation}
\mathrm{Air}[\%] = \left(1 - \frac{\bar{\rho}_{\mathrm{mix}} - \rho_{\mathrm{air}}}{\rho_{\mathrm{water}} - \rho_{\mathrm{air}}} \right)\cdot 100
\end{equation}

Fig.~\ref{fig:mixing} shows the temporal evolution of the air fraction. After impact with the wall, during the run-up phase, a progressive aeration process occurs. This phenomenon becomes more pronounced following jet breakup, reaching values of up to $Air=65\%$  at $T = 4.88$.

During run-down, a deaeration process is observed: the fallback of the jet and the intensification of vortical motions reduce the air fraction in the mixture, gradually returning to the initial equilibrium conditions. This dual mechanism of aeration during run-up and deaeration during run-down is consistent with the experimental observations and will be discussed in detail in Section~\ref{sec:exp_obs}.

After impact ($T = 2.52$), the ascending jet undergoes an intense dissipative process, amplified by the action of the clockwise corner vortex, as reported by \cite{tan2023experimental}. When the jet experiences a breakdown at \( T = 3.94 \), the degree of aeration increases, enhancing its upward deceleration. This process ultimately results in a local flow reversal, which in turn promotes the premature development of a counterclockwise vortex. Maximum aeration occurs when the vortex is fully developed ($T = 4.88$). With the onset of the run-down, the fallback of the main jet further intensifies mixing, while the air fraction progressively decreases until the end of the simulation ($T = 8.00$).
The comparable timescales of the aeration and deaeration processes indicate that the jet does not remain intact during its rise and fall, nor does it maintain a sharp interface. Instead, jet fragmentation and intense water-air mixing trigger an early deaeration process that occurs well before the full development of the run-down (i.e., before the main jet impacts the underlying incoming flow). This aspect cannot be neglected when evaluating the effect of air entrainment on the prediction of the second pressure peak, recently investigated by \cite{rozki2025physical}, as will be discussed in Section ~\ref{sec:press_down}.

\subsubsection{Pressure histories}
\label{sec:press}

Pressure histories obtained with the TSLB model were extracted at the sensor locations installed on the vertical impact wall, as shown in Fig.~\ref{fig:schema}. Since the spatial extent of the pressure sensor exceeds the numerical grid spacing, an averaging procedure has been applied to the numerical results over the corresponding sensor area.

\begin{figure}
\centering

\includegraphics[trim={0cm 11cm 18cm 0cm},clip,height=0.21\textwidth]{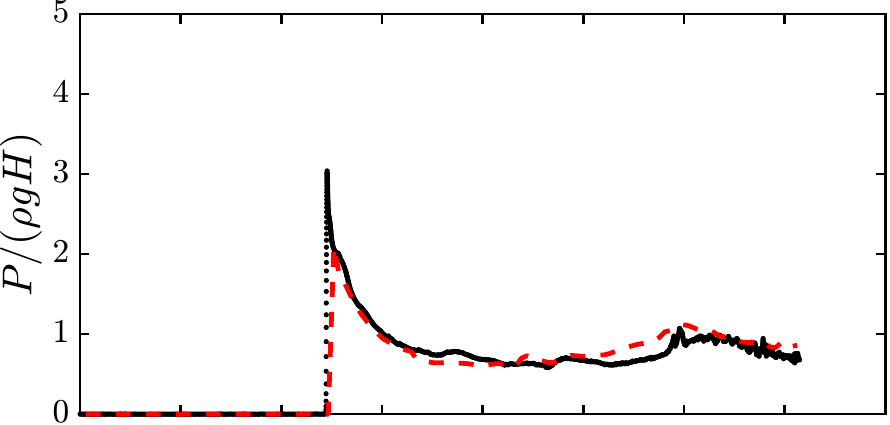}
\includegraphics[trim={19.1cm 11.cm 0cm 0cm},clip,height=0.21\textwidth]{figures/Press_Lob.pdf}

\includegraphics[trim={0cm 0cm 17.95cm 9.4cm},clip,height=0.257\textwidth]{figures/Press_Lob.pdf}
\includegraphics[trim={19.15cm 0cm 0cm 9.4cm},clip,height=0.257\textwidth]{figures/Press_Lob.pdf}

\caption{Temporal evolution of the pressure measured at sensors located along the vertical wall: Sensor 1 (top left), Sensor 2 (top right), Sensor 3
(bottom left), and Sensor 4 (bottom right). Comparison between TSLB numerical results and experimental data \cite{lobovsky2014experimental}.}
\label{fig:pressure}
\end{figure}

A comparison with experimental data, presented in Fig.~\ref{fig:pressure} in terms of dimensionless pressure $p/(\rho g H)$, shows the ability of the model to reproduce the temporal evolution of the pressure signals. In particular, slight differences in the prediction of the first pressure peak can be observed for the sensors located near the base of the vertical wall (Sensors 1 and 2). Similar discrepancies in the estimation of the initial peak, at sensors positioned near the base of the wall, have also been reported by several authors for the same benchmark case \cite{rozki2025physical},
\cite{peng2021numerical}, \cite{garoosi2022experimental}. This suggests that the first pressure peak is strongly influenced by the local impact dynamics and is therefore particularly challenging to predict accurately, as discussed in details in the next section.
Conversely, for Sensors~3 and~4, these differences relative to the experimental measurements become progressively less significant, as they are located further away from the region affected by the impact of the wave front against the vertical wall.
These considerations are consistent with the results of \cite{mokrani2016conditions}, who demonstrated that the calculated pressure peaks exhibit different behaviors depending on whether they are located above or below the stagnation point.
Finally, it is worth noting that the incompressible 3D model adopted in the present simulations accurately predicts the second pressure peak, suggesting that air-compressibility effects play a secondary role during this stage of the impact process, as will be discussed in Section~\ref{sec:press_down}.

\subsubsection{Influence of impact dynamics on first pressure peak uncertainty}
\label{subsecB}
The uncertainty in the prediction of pressure peaks at the base of the vertical impact wall, induced by a dam-break flow, represents a well-recognized problem investigated in various experimental studies \cite{aureli2015experimental}, \cite{xie2022water}, \cite{martinez2026experimental}. This issue was also analyzed in the work by \cite{lobovsky2014experimental}, who conducted an extensive experimental campaign by repeating the same test 100 times under identical initial conditions.

Their statistical analysis revealed that Sensor 1 (see Fig.~\ref{fig:schema}), located 3 mm above the bottom and therefore directly impacted by the incoming wave front, exhibited the largest variability in the recorded pressure peaks. Furthermore, the authors showed that this variability was not significantly correlated with the gate opening speed. 

\begin{figure}
\centering
\includegraphics[trim={0cm 0cm 0cm 0cm}, clip, width=0.5\textwidth]{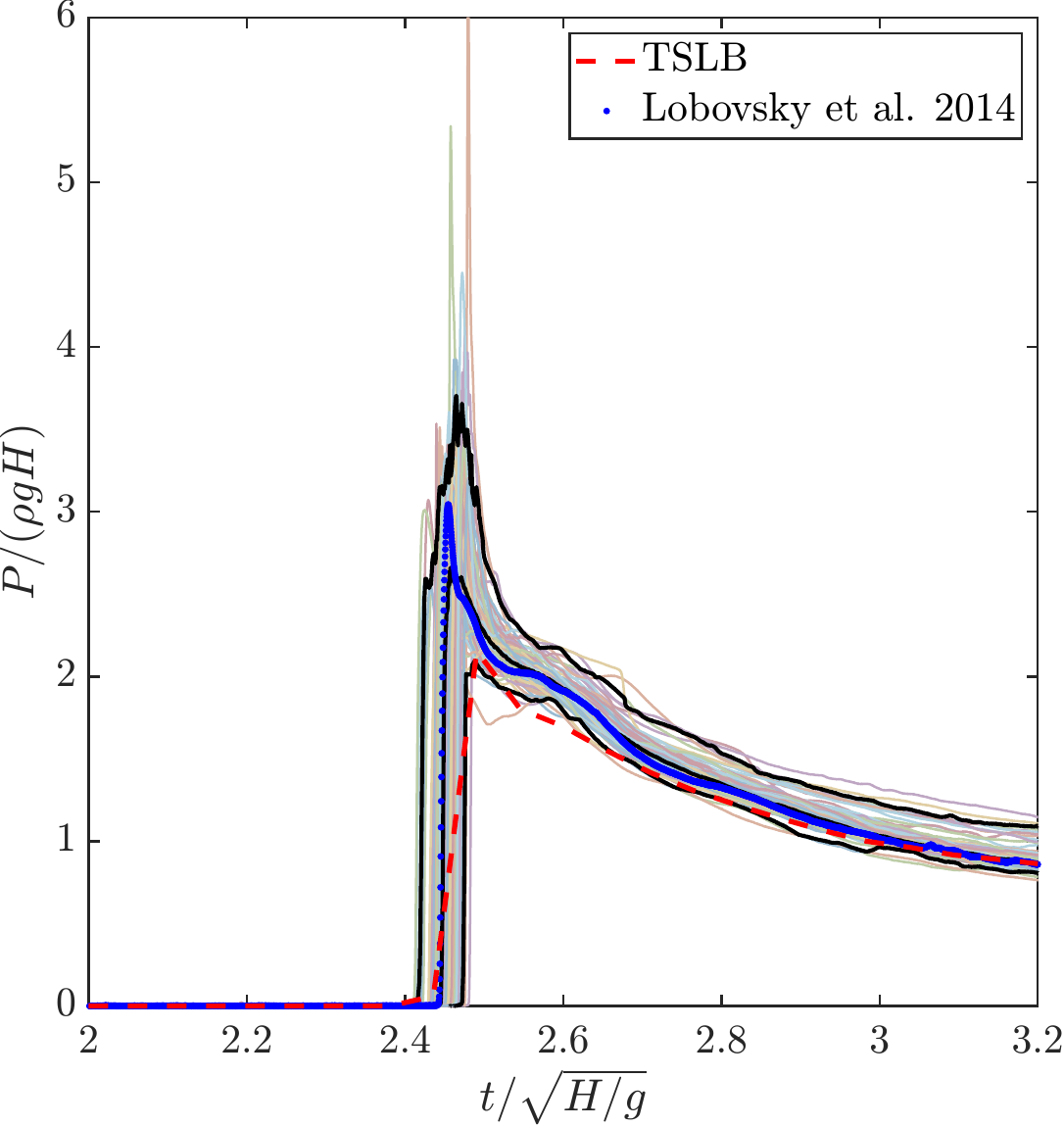}
\caption{Pressure peaks for Sensor 1 from 100 experimental tests by \cite{lobovsky2014experimental}. The solid black line shows the median, with the lower and upper bounds representing the $2.5\%$ and $97.5\%$ percentiles. Blue markers depict a typical experimental pressure history compared against the TSLB numerical solution (dashed red line) (see Fig. \ref{fig:pressure}).}
\label{fig:uncertainty}
\end{figure}

Fig.\ref{fig:uncertainty} shows the interval bounded by the $2.5^{th}$ and $97.5^{th}$ percentiles of the pressure peaks measured at Sensor 1. The TSLB solution falls within this range and lies close to its lower bound, corresponding to the $2.5^{th}$ percentile. This comparison highlights the intrinsically uncertain nature of the first pressure peak, whose magnitude is strongly influenced by localized impact dynamics occurring near the bottom-wall corner.

\subsection{Dam-break flow against a vertical wall (Case II)}
\label{subsec1}

\begin{figure}
\centering
\includegraphics[trim={1.2cm 2.5cm 2.5cm 2.5cm}, clip, width=0.6\textwidth]{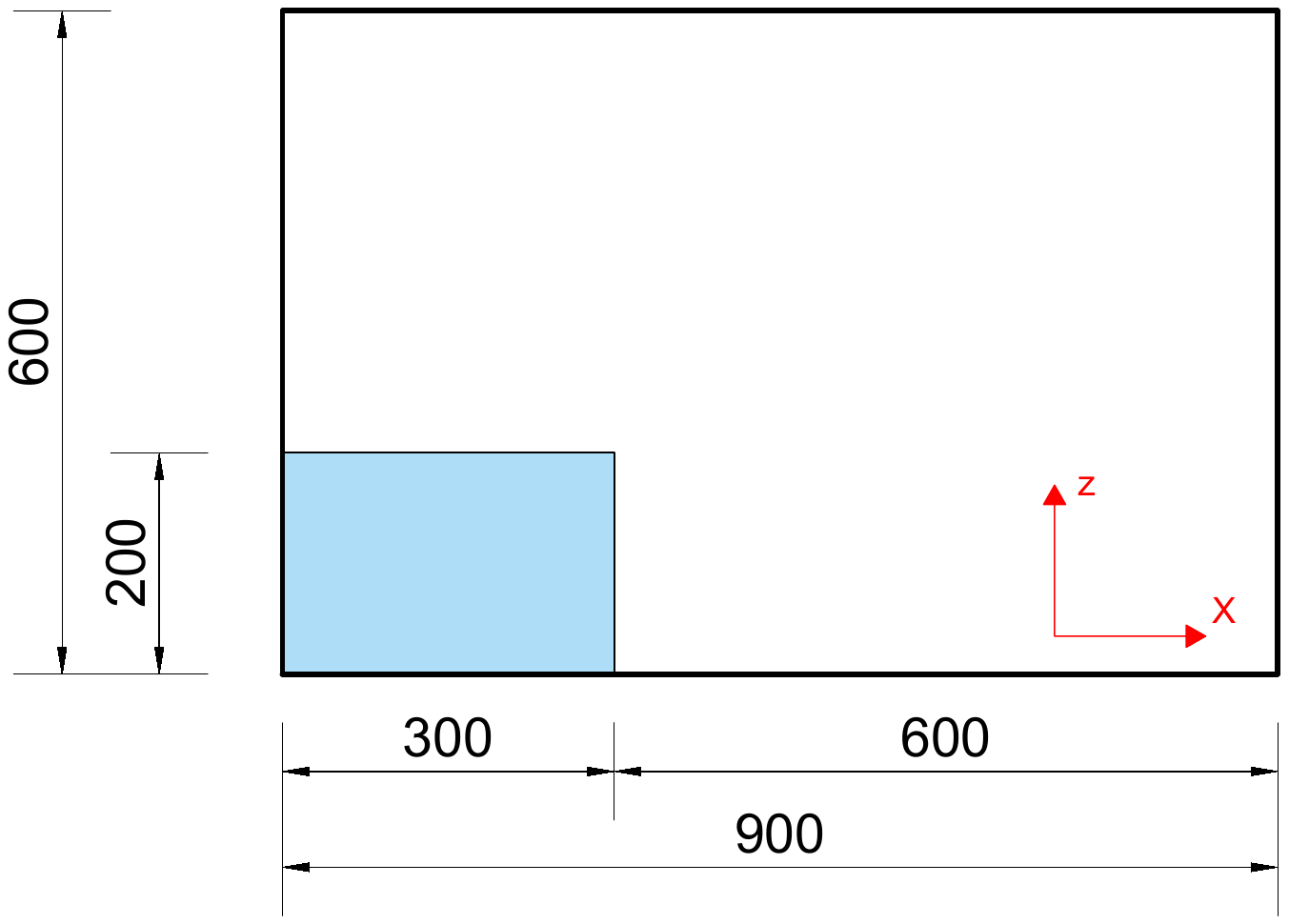}
\caption{Schematic representation of the experimental domain (Case II), dimensions in mm \cite{tan2023experimental}.}
\label{fig:Tan}
\end{figure}

\begin{figure}

\centering
\includegraphics[trim={4.3cm 7.5cm 0.2cm 0.2cm}, clip, width=1\textwidth]{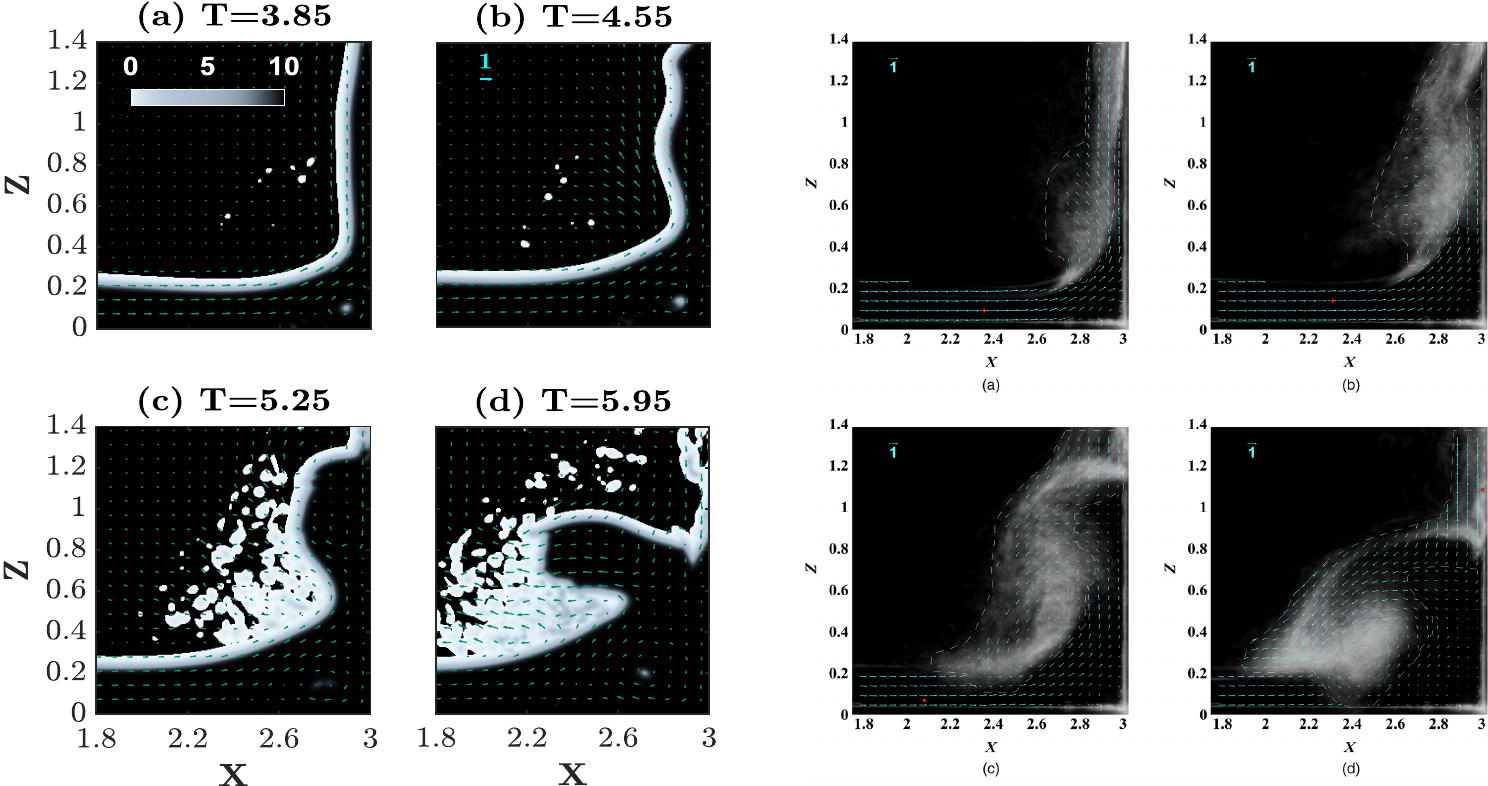}

\caption{Comparison of the free-surface profiles at different times: (a) $T = 3.85$, (b) $T = 4.55$, (c) $T = 5.25$, and (d) $T = 5.95$. Lattice density field obtained with the TSLB model (left) and experimental results from \cite{tan2023experimental} (right). For the numerical results, vectors are drawn every $23$ lattice nodes.}

\label{fig:free_surf}
\end{figure}

The second case study considers the experimental investigation of a dam-break flow over a dry bed impacting a vertical wall, carried out by \cite{tan2023experimental}, as shown in Fig.~\ref{fig:Tan}. The physical domain has a length of 0.9~m, a height of 0.6~m, and a width of 0.25~m. The initial water column is 0.2~m high and 0.3~m long.

Although the dynamics investigated by this experiment is analogous to the one described in Section \ref{subsecA}, additional flow velocity measurements were carried out in the work of \cite{tan2023experimental}. To describe the flow kinematics, the Particle Image Velocimetry (PIV) technique was employed in the non-aerated region, while the Bubble Image Velocimetry (BIV) technique was used to capture the impact of the turbulent flow in the highly aerated region.

Fig.~\ref{fig:free_surf} shows a qualitative comparison of the free-surface profiles at different time instants between the experimental observations and the numerical model.

The TSLB modeling approach is able to effectively reproduce the air--water mixing, which is a key feature for accurately describing the flow kinematics in the post-impact phases.

\subsubsection{Velocity magnitude mapping}
The velocity magnitude distribution, compared with experimental data, was obtained by nondimensionalizing the horizontal and vertical components as $U = u/\sqrt{gH}$ and $W = w/\sqrt{gH}$. The dimensionless velocity magnitude in the $X$--$Z$ plane is defined as $V = \sqrt{U^2 + W^2}$. The dimensionless coordinates are given by $X = x/H$ and $Z = z/H$.

\begin{figure}

\centering
\includegraphics[trim={2.8cm 5.8cm 0.2cm 1.3cm}, clip, width=1\textwidth]{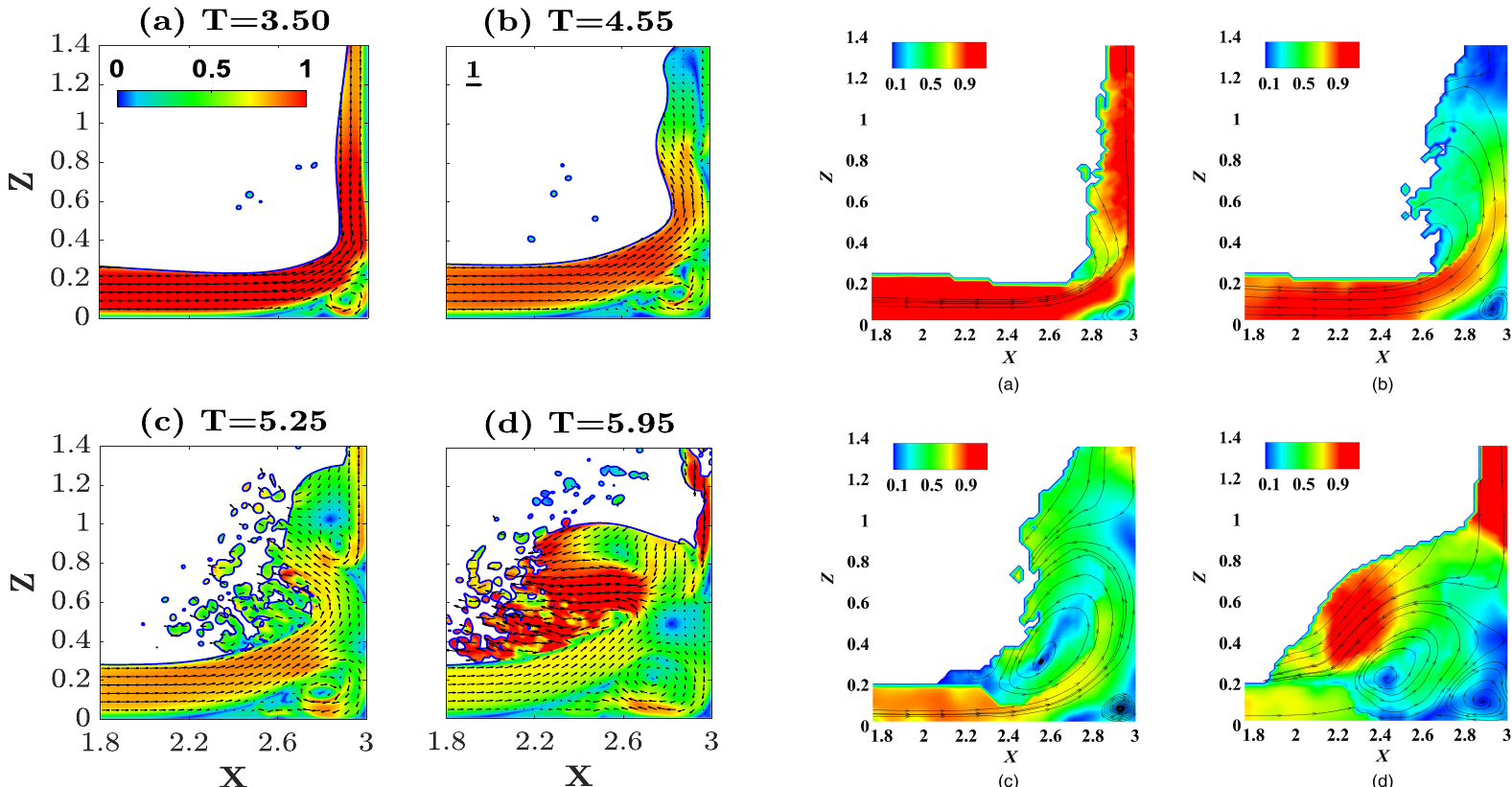}

\caption{Comparison of the velocity magnitude mapping at different times: (a) $T = 3.50$, (b) $T = 4.55$, (c) $T = 5.25$, and (d) $T = 5.95$. Nondimensional values obtained with the TSLB model (left) and experimental results from \cite{tan2023experimental} (right). For the numerical results, vectors are drawn every $18$ lattice nodes.}

\label{fig:vel_map}
\end{figure}

\begin{figure}

\centering
\includegraphics[trim={0cm 0cm 0cm 0cm}, clip, width=0.8\textwidth]{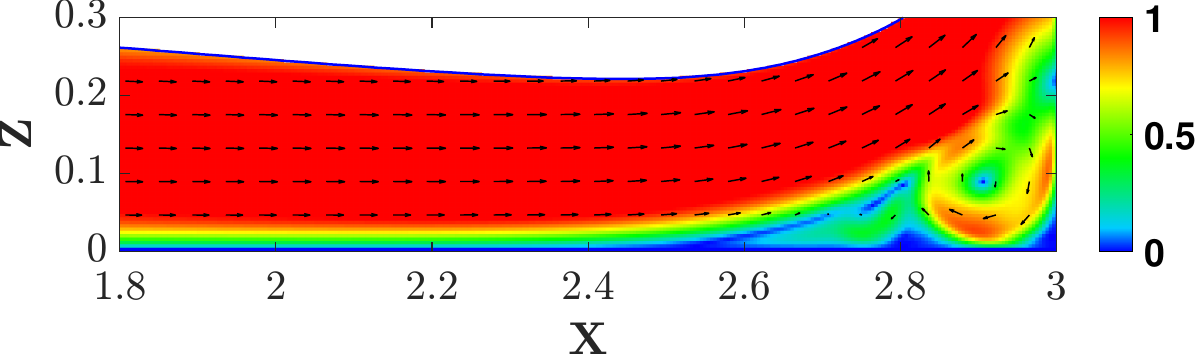}

\caption{Detailed view of the dimensionless velocity gradients within the boundary layer, generated by the no-slip condition applied to the bottom and vertical wall. The results are obtained through the high-resolution DNS of the TSLB model. Vectors are drawn every $18$ lattice nodes.}

\label{fig:bound_layer}
\end{figure}

Fig.~\ref{fig:vel_map} shows that the TSLB model reproduces the temporal evolution of the velocity field in good agreement with experimental measurements. The fine spatial discretization further enables the resolution of boundary layer velocity gradients not accessible experimentally, as visible in the detail shown in Fig.~\ref{fig:bound_layer} 

At $T = 3.50$, the flow enters the run-up phase, with nearly uniform velocity profiles in the uprushing jet, consistent with the solid-body motion assumption by \cite{ko2018splash}. At $T = 4.55$, jet breakup and aeration induce dissipative effects, leading to deviations from uniformity. At $T = 5.25$, a fully developed counterclockwise vortex involving both the aerated jet and the incoming flow is observed. At $T = 5.95$, during the run-down phase, the model captures both the velocity increase in the jet impact region and the nearly uniform velocity of the aerated spray along the wall.

The clockwise vortex develops at the bottom corner due to boundary layer effects along the bottom and the wall. Initially ($T = 3.50$), the vortex is small and exhibits intense vorticity, associated with thin boundary layers and strong velocity gradients. During run-up ($T = 4.55$, $T = 5.25$), boundary layer thickening and Reynolds number reduction lead to an increase in vortex size and a decrease in intensity. This trend persists during run-down ($T = 5.95$), with further vortex expansion and weakening, consistent with reduced velocity gradients.

\begin{figure}

\centering
\includegraphics[trim={0cm 0cm 0cm 0cm}, clip, width=0.9\textwidth]{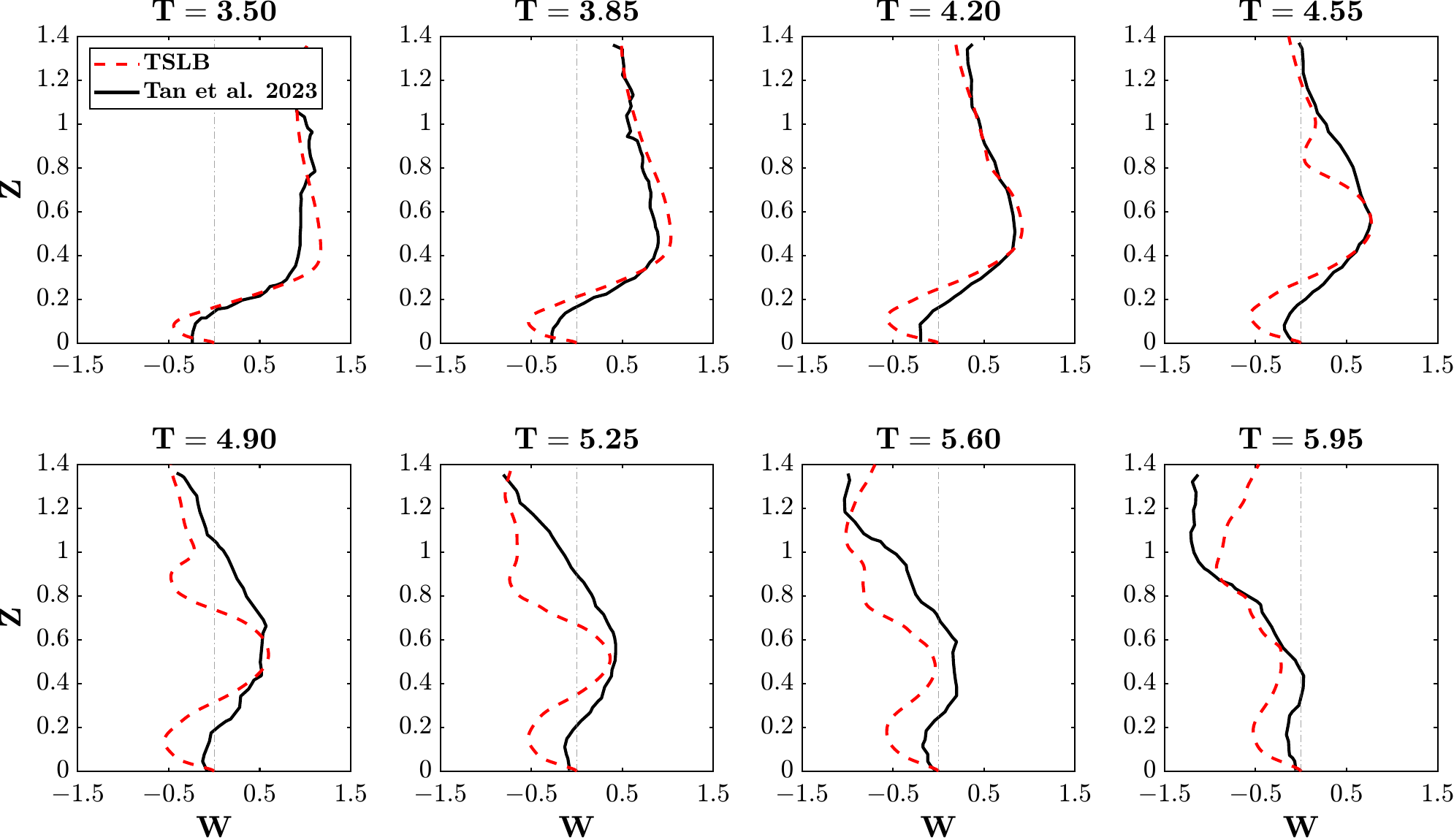}

\caption{Comparison of vertical velocity profiles along the impact wall between the TSLB numerical results and the experimental data of \cite{tan2023experimental}. The vertical dashed gray line marks the zero-velocity elevation (W=0).}

\label{fig:vel_prof}
\end{figure}

\subsubsection{Vertical velocity profile}

The comparison of the vertical velocity profiles, $W = w/\sqrt{gH}$, along the impact wall at different time instants is shown in Fig.~\ref{fig:vel_prof}. The numerical results obtained using the TSLB modeling framework should be interpreted as mean profiles, computed from velocity values extracted over multiple sections located at different distances from the vertical wall. Specifically, four sections between 9~mm and 16~mm from the wall were considered, each characterized by a different degree of influence of wall friction effects.

During the run-up phase, a good agreement with the experimental results is observed. At the first analyzed instant ($T = 3.50$), the vertical velocity is nearly uniform for $Z > 0.3$, in agreement with the solid-body motion assumption proposed by \cite{ko2018splash}.

At later times ($T = 3.85$, $T = 4.20$), this uniformity progressively breaks down. At the elevation $Z = 1.4$, the vertical velocity gradually tends toward zero, while remaining relatively higher in the interval $0.3 < Z < 0.6$. This behavior can be attributed to the incoming flow that persists during this phase. Even at these elevations, however, the vertical velocity exhibits a gradual reduction over time, as evidenced at $T = 4.55$.
During the run-down phase, at $T = 4.90$ and $T = 5.25$, the vertical velocity at $Z = 1.4$ becomes negative, indicating a downward reversal of the flow streamlines. Similarly, in the interval $0.3 < Z < 0.6$, the velocities progressively decrease as a result of the deceleration of the incoming flow. Finally, at $T = 5.60$ and $T = 5.95$, for $Z > 0.9$, the vertical velocity tends to become uniform again along the vertical direction, due to the fallback of the aerated spray along the wall, which is not effectively captured by the modeling approach.

At elevations $Z < 0.3$, the TSLB modeling correctly reproduces the velocity gradients in the vicinity of the bottom. In the initial phase, the boundary layer is thin and, consequently, the velocity gradients are high. As time progresses, however, the boundary layer progressively thickens, leading to a reduction of the vertical velocity gradients.

\section{Discussion}
\label{sec2}

\begin{figure}
\centering
\includegraphics[trim={16.0cm 6.5cm 0cm 0cm}, clip, width=0.7\textwidth]{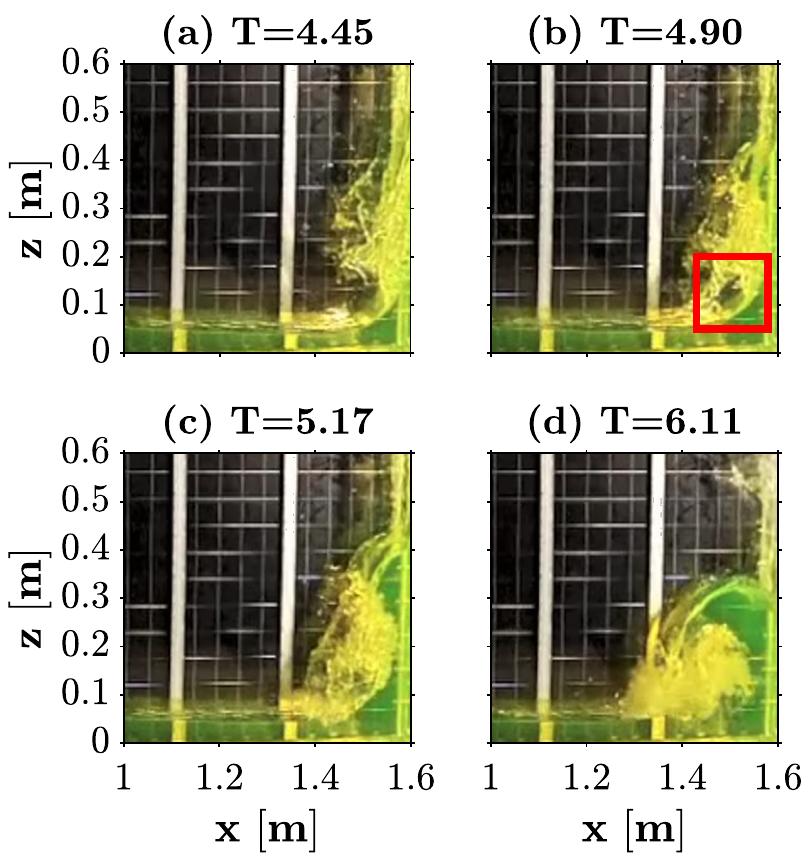}

\caption{Formation of the counterclockwise vortex during the run-up phase of the dam-break flow, with the entrapment of a small air cavity within the vortical region, visible at time $T = 4.90$. Experimental test by \cite{lobovsky2014experimental}.}

\label{fig:vortex}
\end{figure}

\subsection{Premature onset of the counterclockwise vortex}
\label{sec:exp_obs}

Fig.~\ref{fig:vortex} presents additional snapshots of the experimental run reported in Fig.~\ref{fig:profile}, covering the phase from the development of the run-up to the subsequent run-down phase. 

At the instant $T=4.45$, the images show that, following the breakup of the jet during the run-up phase, the fragmented water falls back onto the incoming flow, promoting the initiation of a counterclockwise vortex. 

At $T=4.90$, the entrapment of a small air cavity within the vortical region is also observed, indicating the complete development of the vortex at this stage. The presence of the vortex promotes mixing of the aerated water before the run-down phase fully occurs, as visible at $T=5.17$.  

These observations therefore indicate that the counterclockwise vortex, caused by the downward reversal of the flow lines, is already activated during the run-up phase following the breakup of the upward water jet along the wall. This mechanism, also observed in the work of \cite{tan2023experimental}, provides a possible explanation for why, during the run-down phase, the cavity beneath the main jet is already strongly mixed, as visible at $T=6.11$. In this sense, the counterclockwise vortex appears earlier and is located lower than expected in the absence of jet breakup and its subsequent formation during the run-down phase, as evidenced by the comparison between numerical and experimental flow structures in \cite{rozki2025physical}.

In light of these observations, modeling the water–air mixture is crucial for accurately evaluating the dynamics of the counterclockwise vortex.

\subsection{Comparison between no-slip and free-slip results}
\label{subsec1}

In the literature, many numerical simulations of dam-break flows impacting a vertical wall over a dry bed show that the run-down phase of the main jet over the incoming flow is systematically associated with the entrapment of a macroscopic, smooth, large and well defined air cavity \cite{colagrossi2003numerical}, 
\cite{rozki2025physical},
\cite{garoosi2022experimental}, \cite{yang2020numerical}. This scenario manifests not only in models that implement a free-slip boundary condition, but also in many no-slip cases. This evidence makes it particularly challenging to determine whether the failure to reproduce the complex air–water mixture—which is instead clearly visible in experimental tests during the run-down phase—is the result of an intrinsic limitation of the numerical model or, conversely, is attributable to external physical factors that cannot be replicated in an idealized simulation, such as the gate opening speed or geometric imperfections of the experimental channel.

\begin{figure}

\centering
\includegraphics[trim={4.2cm 4.0cm 0.2cm 1.3cm}, clip, width=1.\textwidth]{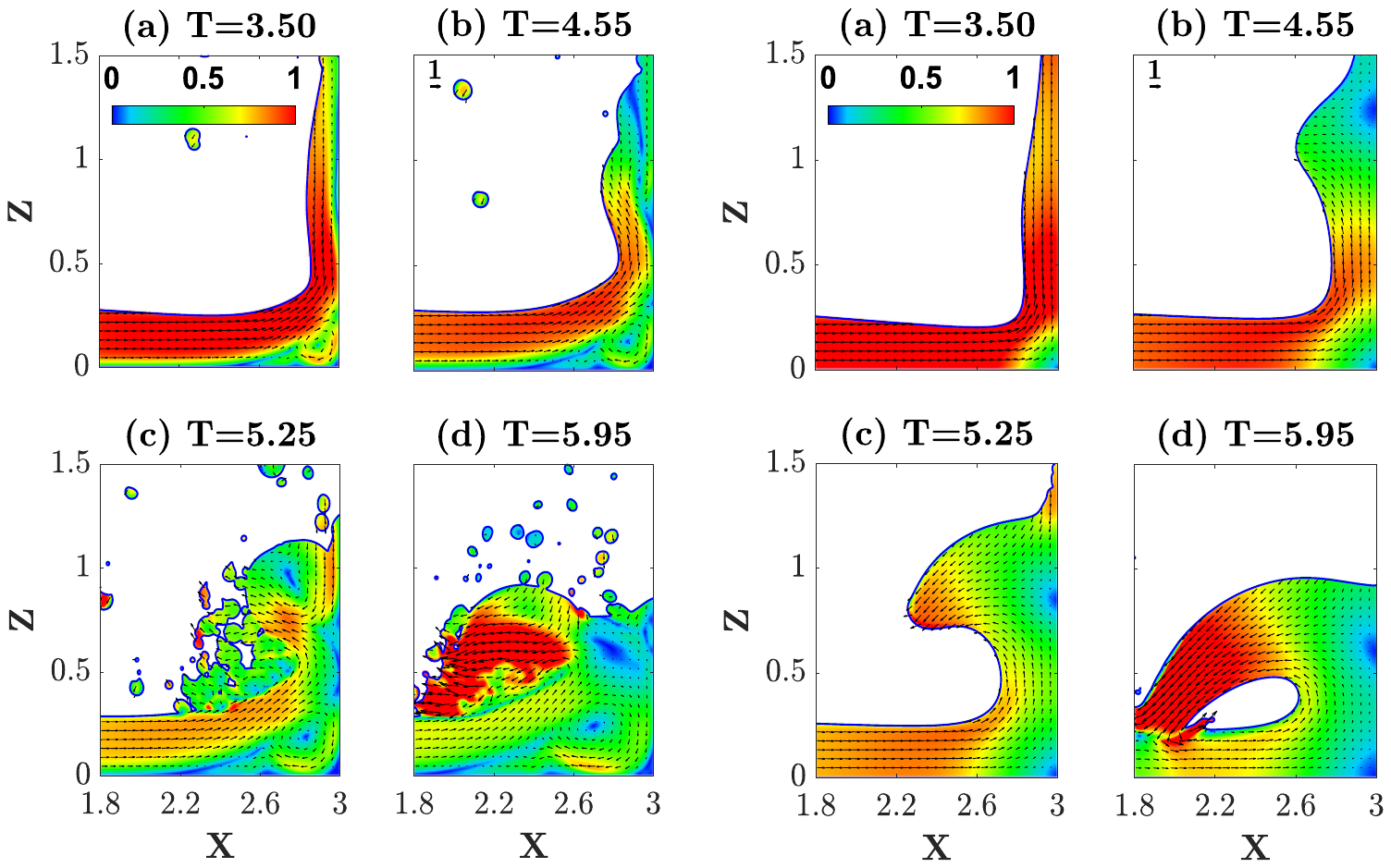}

\caption{Comparison between no-slip (left) and free-slip (right) conditions of the velocity magnitude maps at the same time instants. Vectors are drawn every $18$ lattice nodes.}

\label{fig:vel_slip}
\end{figure}

To isolate and analyze this issue, the numerical model configuration employed in the benchmark tests of \cite{tan2023experimental} was implemented herein to compare two simulations characterized by identical geometric similarity and the same Reynolds number ($Re = 2.2 \times 10^3$), differing solely in the boundary condition applied at both the bottom and vertical wall. Fig.~\ref{fig:vel_slip} presents a comparison, in terms of dimensionless velocity magnitude fields at various time instants, of the dam-break flow under no-slip and free-slip conditions. Despite all other computational parameters being identical, the two configurations exhibit markedly different flow evolutions.

Enforcing the no-slip condition along the walls generates strong velocity gradients and intense shear stresses that amplify perturbations at the air–water interface, triggering local hydrodynamic instabilities that lead to jet breakup and intense two-phase mixing during the post-impact phase. Conversely, the absence of tangential stresses in the free-slip configuration suppresses vorticity generation at the wall from the very beginning. Consequently, the water jet maintains a nearly uniform vertical velocity profile over a longer time interval, remaining closer to the solid-body motion assumption of \cite{ko2018splash}, and promoting the formation of a large air cavity during the run-down phase.

This comparative analysis therefore suggests that the macroscopic air cavity described in the literature, and here reproduced by the free-slip simulation (right panels of Fig.~\ref{fig:vel_slip}),  may not necessarily be linked to unavoidable discrepancies with experimental reality, but may instead be the result of numerical artifacts arising from model limitations.

This observation is particularly relevant in Reynolds Averaged Navier Stokes (RANS) simulations. When the near-wall region is under-resolved, the first computational point may lie outside the range of validity of the employed wall functions \citep{wilcox1993turbulence}. Under these conditions, wall functions are applied outside their range of validity and are no longer capable of correctly reconstructing the steep velocity gradients within the boundary layer.

Consequently, despite formally enforcing a no-slip boundary condition, such numerical solutions may dynamically approach a free-slip-like behavior.

The results obtained in the present work demonstrate that an adequate resolution of the boundary layer is a fundamental requirement for the accurate simulation of dam-break flows. In the case investigated here, a DNS with a spatial resolution sufficiently fine to resolve near-wall velocity gradients successfully reproduces both the jet breakup process and the subsequent air–water mixing observed experimentally.
This conclusion is further supported by the work of \cite{del2025turbulent}, who showed through three-dimensional simulations that turbulent dam-break waves develop a well-defined two-layer structure, including a near-wall logarithmic sublayer. Their study highlighted that accurately resolving this region is essential for correctly capturing flow resistance and bed shear stresses.

\begin{figure}

\centering
\includegraphics[trim={0cm 6.8cm 3.7cm 0.2cm}, clip, width=1.\textwidth]{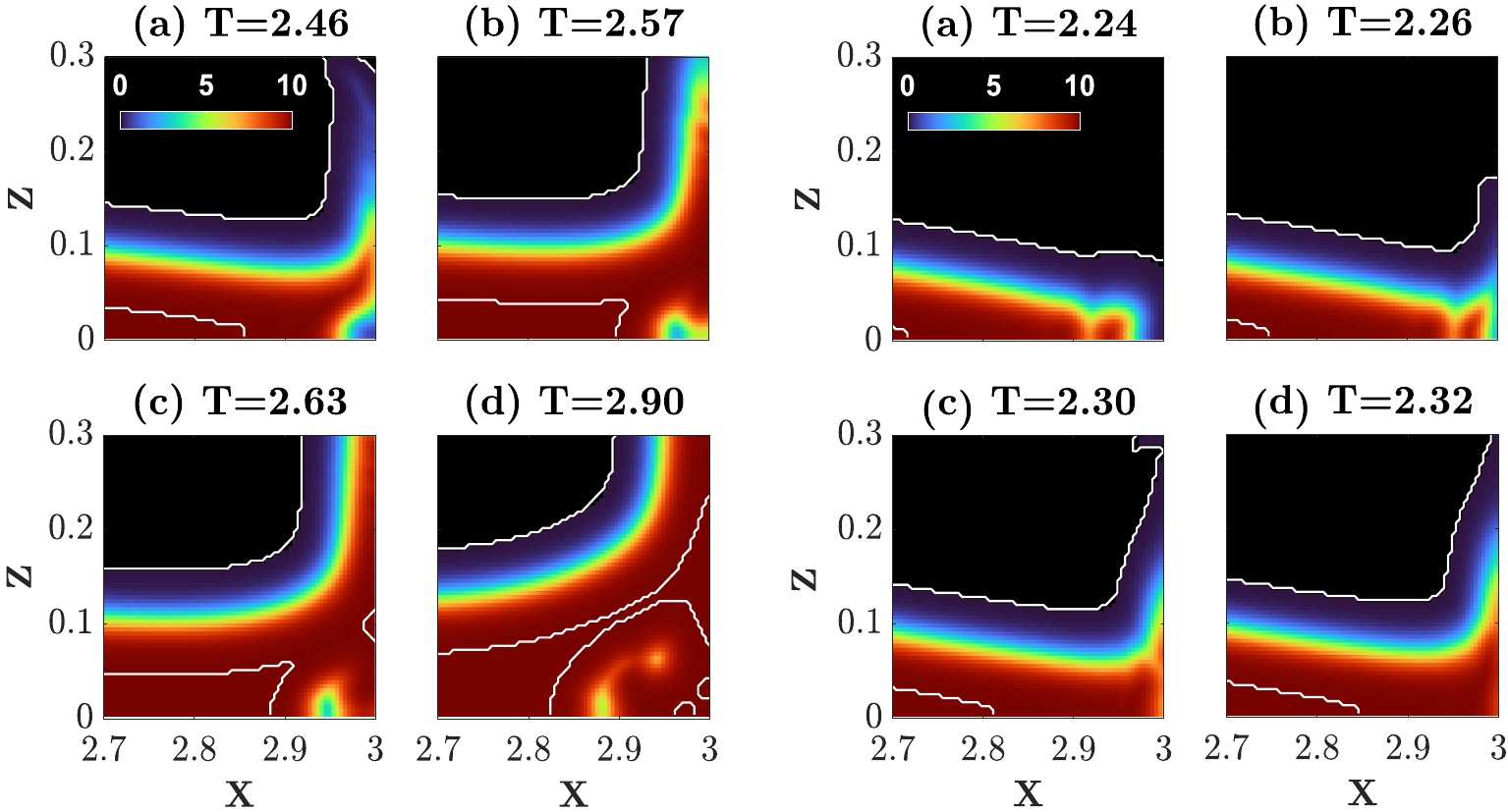}

\caption{TSLB numerical density maps: comparison between no-slip (left) and free-slip (right) conditions at the time instants corresponding to the impact of the dam-break wave against the vertical wall.}

\label{fig:impact}
\end{figure}

\begin{figure}

\centering
\includegraphics[trim={0cm 0cm 0cm 0cm}, clip, width=0.75\textwidth]{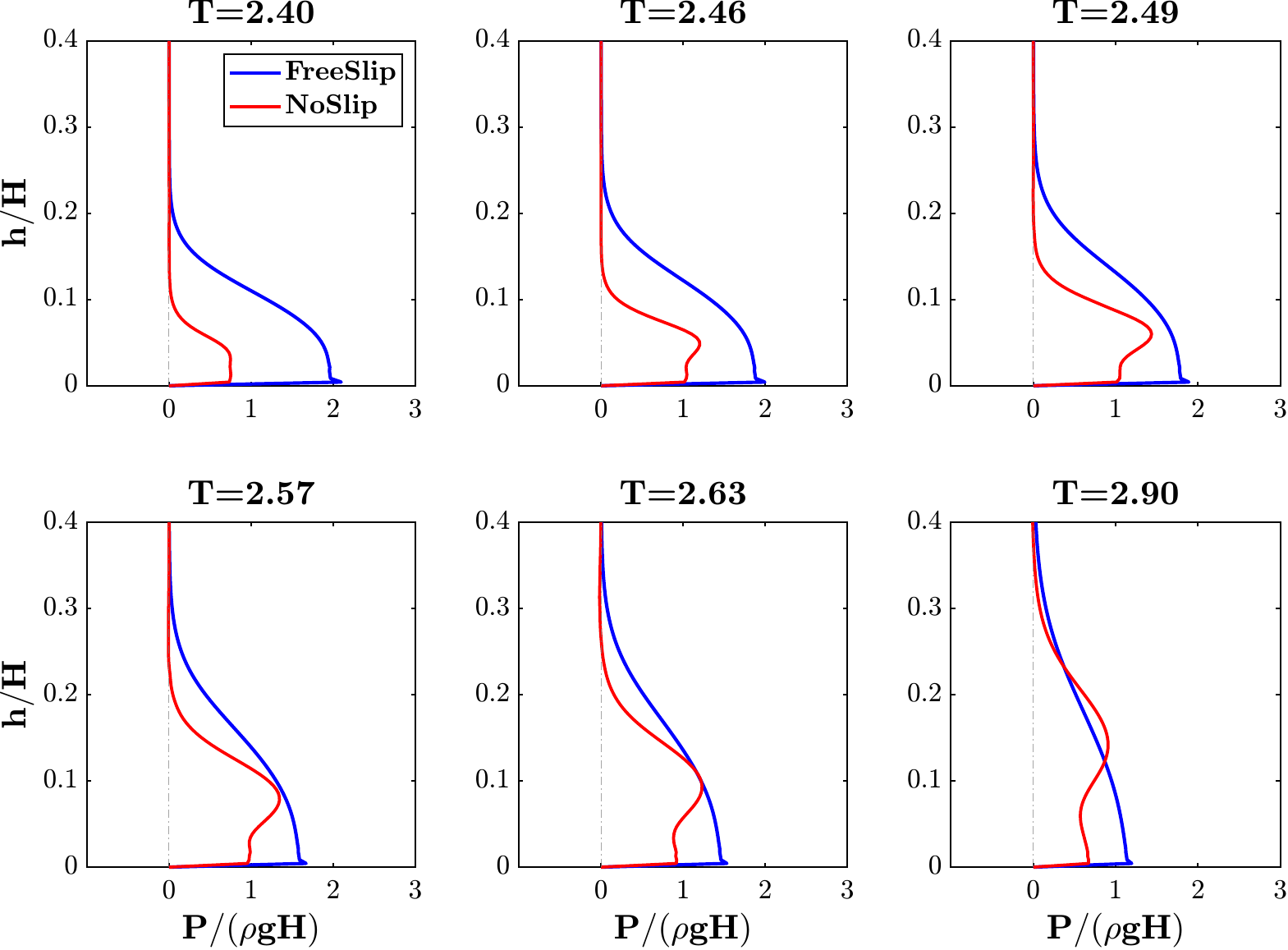}

\caption{Pressure distribution along the vertical wall during the initial stages of wave tip impact: comparison between no-slip and free-slip conditions.}

\label{fig:confronto_press}
\end{figure}

\subsubsection{Shear-induced aeration and jet deflection}
\label{sec:small_vort}

Fig. \ref{fig:impact} compares the impact dynamics yielded by a no-slip and free-slip simulations. The comparison highlights that the clockwise vortex developing at the bottom corner is absent under free-slip conditions. The absence of the boundary layer prevents the vortex formation and, consequently, no air entrainment occurs in the near-bottom region.

Moreover, during the impact phase, the corner vortex plays a crucial role in promoting the upward deflection of the wave front. As a result, the location of maximum pressure along the vertical wall does not occur exactly at the wall base but above the small clockwise corner vortex. Conversely, this characteristic is not predicted with the free-slip model, as shown in Fig.~\ref{fig:confronto_press}.

Further evidence of these distinct impact dynamics is confirmed by comparisons between the experimental pressure distributions measured by \cite{kihara2015large}  and the theoretical predictions of \cite{cumberbatch1960impact}, as well as by the numerical results of \cite{mokrani2016conditions}. In particular, the latter demonstrated that, when frictionless
boundary conditions are adopted, pressure peaks near the impact point tend to diverge as the spatial resolution increases, unlike what occurs when the turbulent boundary layer is adequately resolved.

The high-resolution DNS presented in this article therefore allow for an accurate reproduction of the local impact dynamics, providing important insights into the physical origin underlying the uncertain nature of the first pressure peak (as seen in Section \ref{subsecB}) and into the central role played by the boundary layer and the corner vortex in the generation of impulsive loads on vertical walls.

Once air is entrained within the vortex under no-slip conditions, an intense air–water mixing process is triggered near the corner prior to jet breakup during the run-up phase. This mechanism provides a possible explanation for the high turbulence intensity measured near the bottom by \cite{tan2023experimental}. It follows that the process of air entrainment is not limited to run-up and run down phases: indeed, the near-bottom region, through the associated corner vortex dynamics, serves as a significant source of finely mixed air for the upward-travelling flow.
 
This work therefore quantitatively supports the conclusions of \cite{mokrani2016conditions}, highlighting the prominent role of air–water interaction, as well as boundary conditions, in accurately reproducing the shape of the wave front tip, which plays a fundamental role in the impact process.

\begin{figure}
\centering

\includegraphics[trim={0cm 8cm 7cm 0cm},clip,height=0.36\textwidth]{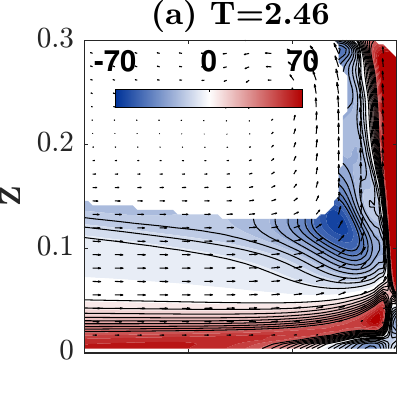}
\includegraphics[trim={7.8cm 8cm 0cm 0cm},clip,height=0.36\textwidth]{figures/vorticity_NoSlip_Tan.pdf}

\includegraphics[trim={0cm 0cm 7cm 7.5cm},clip,height=0.385\textwidth]{figures/vorticity_NoSlip_Tan.pdf}
\includegraphics[trim={7.8cm 0cm 0cm 7.5cm},clip,height=0.385\textwidth]{figures/vorticity_NoSlip_Tan.pdf}

\caption{Nondimensional vorticity maps at the time instants corresponding to the impact of the dam-break wave against the vertical wall, under no-slip boundary condition. Vectors are drawn every $5$ lattice nodes.}
\label{fig:vorticity}
\end{figure}

The inspection of the vorticity distribution further supports the above analysis. Fig.~\ref{fig:vorticity} shows the dimensionless vorticity field during the impact phase under no-slip conditions. The vorticity is computed from the dimensionless velocity components $U = u/\sqrt{gH}$ and $W = w/\sqrt{gH}$ as
\begin{equation}
\omega_y = \frac{\partial U}{\partial Z} - \frac{\partial W}{\partial X}.
\end{equation}

\begin{figure}
\centering
\includegraphics[trim={0.0cm 0cm 0cm 0cm}, clip, width=0.7\textwidth]{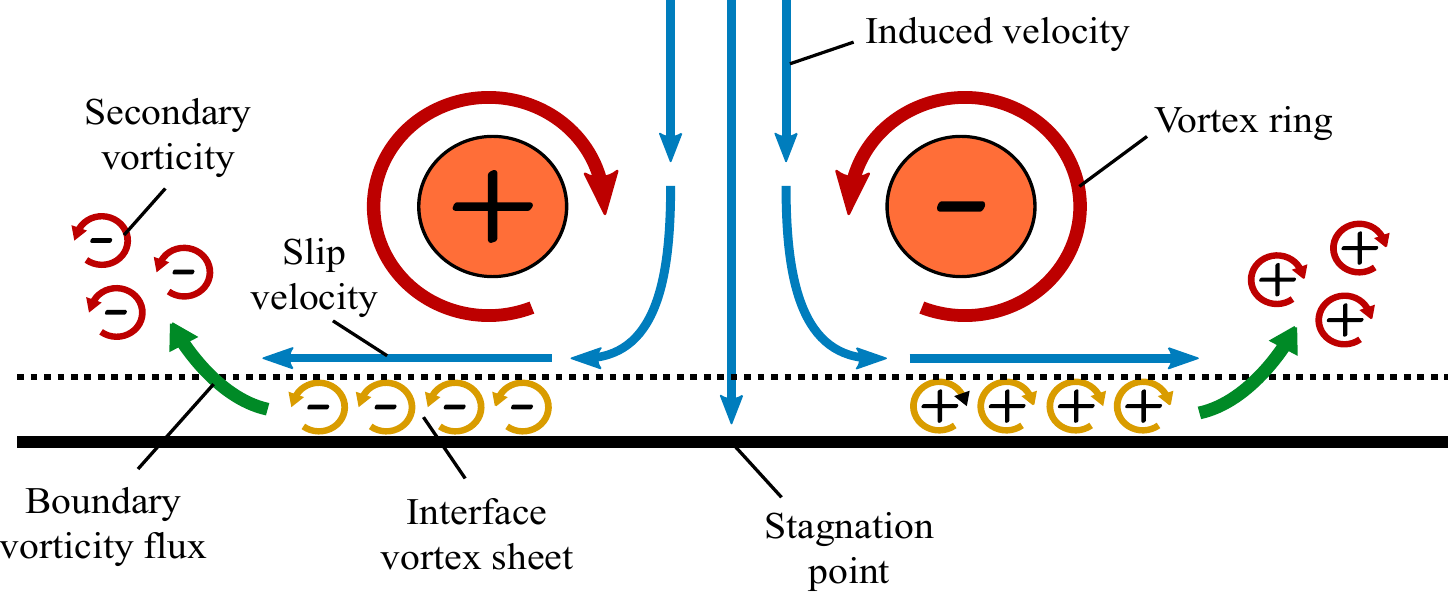}

\caption{Mechanism of generation of secondary counterclockwise vorticity at the bottom--wall corner. Reprinted by \cite{terrington2024vorticity} under the terms of the Creative Commons CC-BY license.}

\label{fig:terrington} 
\end{figure}

A strong clockwise vorticity is observed along the bottom, driven by velocity gradients within the boundary layer. At the bottom–wall corner, the no-slip condition on both the bottom and the wall promotes the formation of a clockwise corner vortex. During the impact, while part of the fluid rises along the wall, another portion is deflected downward (as shown in Fig.~\ref{fig:impact}). This local flow reversal leads to the generation of secondary counterclockwise vorticity, also attributable to velocity gradients near the wall. A detailed study on the generation of secondary vorticity is provided by \cite{terrington2024vorticity} and schematically illustrated in Fig.~\ref{fig:terrington}.

As a result, while the primary vortex maintains a clockwise circulation that progressively expands and weakens, the internal air--water mixing is dominated by an intense counterclockwise circulation.

\begin{figure}
\centering
\includegraphics[trim={0.0cm 0cm 0cm 0cm}, clip, width=0.7\textwidth]{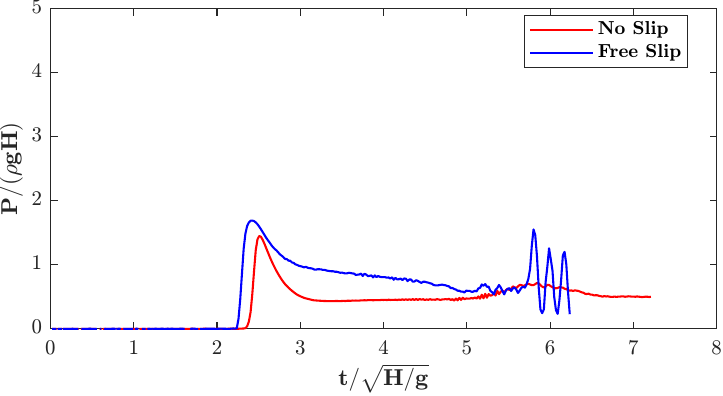}

\caption{Temporal distributions of pressure in the two different conditions of no-slip and free-slip, measured at a distance of 15 mm from the bottom, along the vertical impact wall.}

\label{fig:prof_press}
\end{figure}

\subsubsection{Air entrainment effect on pressure history}
\label{sec:press_down} 

The pressure response was analyzed using an ideal sensor positioned on the vertical wall at an elevation of 15 mm from the bottom. A comparison of the temporal pressure distributions (Fig.~\ref{fig:prof_press}) reveals that the free-slip condition yields a more intense and temporally advanced primary peak compared to the no-slip case. This discrepancy is attributable to the absence of viscous dissipation at the boundary, which entails a faster front advancement and results in an earlier and more energetic impact. Furthermore, under free-slip conditions, the post-impact residual pressure remains higher, due to the lack of damping and local aeration that would otherwise be generated by the corner vortex.

The differences between the two configurations become more pronounced at the second peak. The no-slip condition produces a smoother profile, free of the spurious peaks observed in \cite{rozki2025physical}, showing better agreement with experimental evidence. Conversely, the free-slip simulation induces marked pressure oscillations along the wall, similar to those reported by \cite{michel2025efficient}.

\begin{figure}

 \centering
\includegraphics[trim={4.2cm 5cm 0.1cm 0.1cm}, clip, width=1\textwidth]{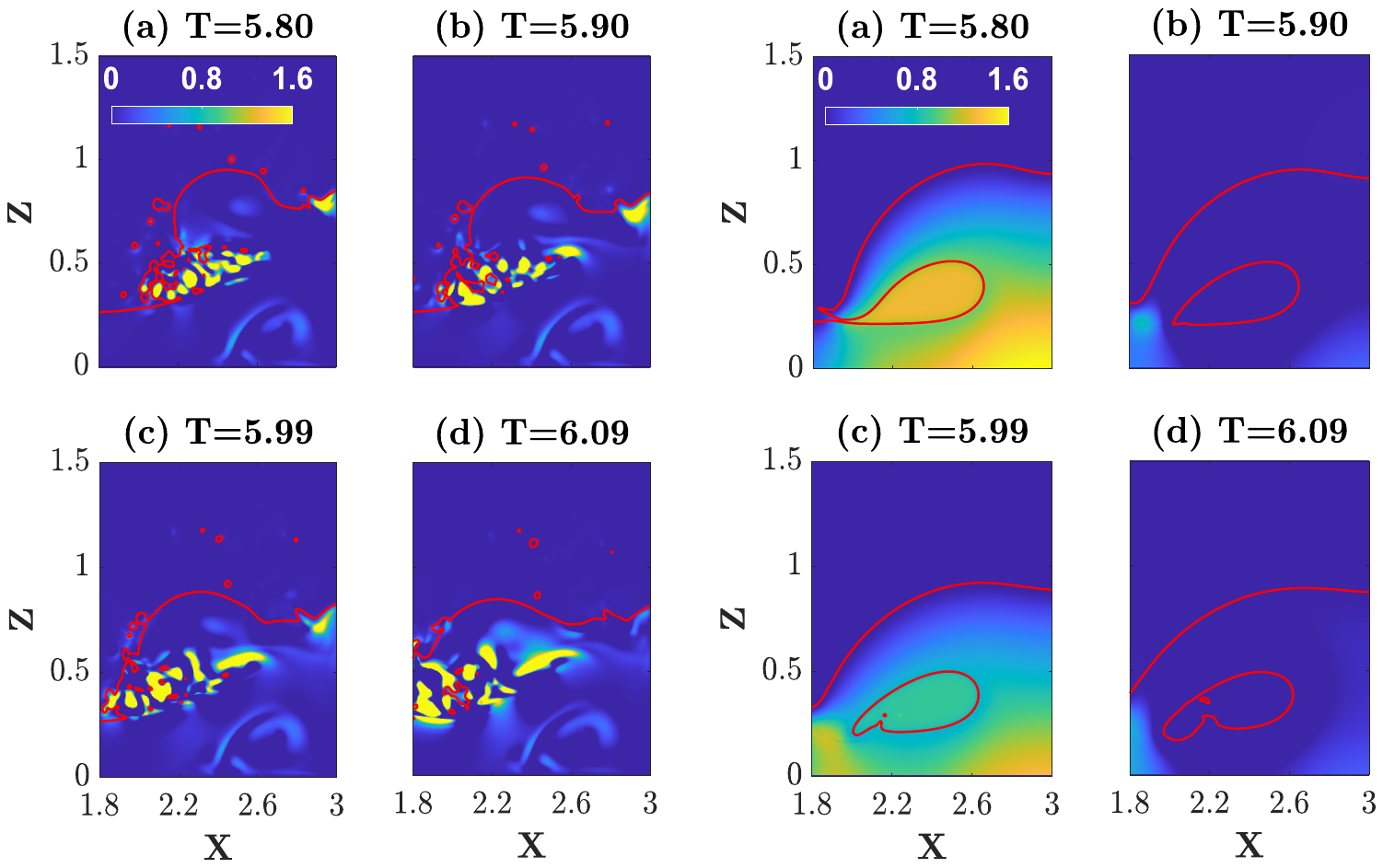}

\caption{Nondimensional pressure field at different time instants during the run-down phase, comparison between the no-slip condition (left) and the free-slip condition (right).}
\label{fig:map_press}
\end{figure}

Temporal evolution of the dimensionless pressure field (Fig.~\ref{fig:map_press}) elucidates the origin of these fluctuations. During the run-down phase, the free-slip configuration exhibits the formation of an intact air cavity beneath the jet, with dynamics resembling a plunging wave (situation in which the ideal fluid assumption remain valid). In this scenario, the volume of entrapped air reaches its maximum at the end of the run-down phase, precisely when the wave crest impacts the incoming free-surface flow. 

Unlike the 2D models adopted by \cite{colagrossi2003numerical} and \cite{rozki2025physical}, where the so-called air cushion effect is associated with compression-expansion cycles of the air cavity, the simulations presented here utilize an incompressible 3D framework. In this case, the air cavity maintains a nearly constant volume; thus, the observed oscillations do not originate from gas compression but are instead driven by the geometrical deformation of the cavity and the dynamics of the water jet. 

Conversely, in the no-slip condition, wall friction triggers jet fragmentation and intense two-phase mixing. In this configuration, the air fails to form a single coherent cavity, instead dispersing into smaller structures embedded in the water. These results highlight that, while an intact and well-defined cavity (free-slip case) allows large pressure fluctuations to arise, jet disruption and subsequent air-water mixing (no-slip case) prevent the formation of a coherent elastic-like dynamics, leading to a drastic attenuation of oscillations.

\section{Conclusions}
\label{sec1}

In this article, a multiphase thread-safe lattice Boltzmann (TSLB) model, coupled with the Allen–Cahn equation for air–water interface capturing, has been employed to perform high-resolution simulations of the impact of a dam-break wave against a vertical wall. After successful validation against experimental data, the model was used to investigate the role of air entrainment in determining the impact dynamics.

The simulations indicate that the often reported large air cavity formed during the run-down phase is a numerical artifact of the model rather than a physical phenomenon.

The main findings are summarized as follows:
\begin{enumerate} 

\item Comparisons between simulations performed at the same Reynolds number ($Re$) show that a large, smooth, coherent air cavity develops during the run-down phase only when a free-slip boundary condition is imposed at the wall. In contrast, wall friction amplifies interfacial perturbations, promotes jet breakup, and accelerates the deaeration process before the main jet impacts the incoming flow.
\item Air compressibility has only a marginal influence on the prediction of the second pressure peak. Although a coherent air cavity (free-slip case) generates pressure oscillations, these are significantly damped when jet breakup and the resulting air–water mixing occur under no-slip conditions. These results suggest that the dam-break impact can be accurately reproduced using an incompressible three-dimensional model without generating spurious pressure peaks.
\item During the impact process, the clockwise vortex developing near the lower corner of the wall governs the upward deflection of the incoming flow. An accurate representation of the no-slip boundary condition is therefore essential. Simulations employing a free-slip condition fail to correctly predict the amount of air entrained by the corner vortex and, therefore, the location of the wave-front impact, leading to inaccurate estimates of the distribution of pressure peaks along the vertical wall.
\end{enumerate}
In summary, the results demonstrate that adequate numerical resolution together with an accurate treatment of no-slip boundary conditions is essential for reliable simulations of dam-break impacts. Correct representation of the boundary layer and of the water–air interface is crucial for reproducing the complex flow physics governing air entrainment and pressure generation.

\subsection*{\textbf{Acknowledgments}}

A.M. and P.P. acknowledge funding from the Italian Government through the PRIN project MOBIOS (Grant No. 2022N4ZNH3, CUP: F53C24001000006).

A.M., P.P. and M.M. acknowledge the computational support from CINECA through the ISCRA B project MIPLAST (IsB31, HP10BZY7BK).

M.M. and P.P. acknowledge funding from the project “Ecosistema dell’Innovazione—Rome Technopole” financed by EU-Next-GenerationEU (CUP: F83B22000040006).


\bibliographystyle{elsarticle-num}
\bibliography{biblio_db}





\end{document}